\documentclass[preprint,11pt,3p,review]{elsarticle}

    \journal{}

    \usepackage{paper-style}
    \usepackage{tensor-operators}

    \usepackage{tikz}
		\usetikzlibrary{arrows.meta}

    \tikzset{
        pics/picStatsTest/.style args={#1/#2}{%
            code={%
                \draw[%
                    densely dotted,
                    {Circle[open,length=2pt]}-{Circle[open,length=2pt]},
                ]   (-0.5cm,0cm)
                    -- node[%
                            midway,
                            fill=white,
                            inner sep=0pt,
                            label={[fill=white,inner sep=2pt]#1:{\small #2}}
                        ]   {$p$}
                    ++(1cm,0cm);
            }
        }
    }

\newglossaryentry{radial}{
    type=hidden,
    name={\ensuremath{r}},
    description={Radial coordinate of the cylindric coordinate system}
}
\newglossaryentry{azimuthal}{
    type=hidden,
    name={\ensuremath{\theta}},
    description={Circunferential coordinate of the cylindric coordinate system}
}
\newglossaryentry{z}{
    type=hidden,
    name={\ensuremath{z}},
    description={Axial coordinate of the cylindric coordinate system}
}

\newglossaryentry{xAxis}{
    type=hidden,
    name={\ensuremath{x}},
    description={%
        First coordinate of the Cartesian system (normally, $x_1$)
    }
}
\newglossaryentry{yAxis}{
    type=hidden,
    name={\ensuremath{y}},
    description={%
        Second coordinate of the Cartesian system (normally, $x_2$)
    }
}
\newglossaryentry{zAxis}{
    type=hidden,
    name={\ensuremath{z}},
    description={%
        Third coordinate of the Cartesian system (normally, $x_3$)
    }
}

\newglossaryentry{average}{
    type=subscript,
    name={\ensuremath{avg}},
    description={Indicates a time and surface averaged variable}
}
\newglossaryentry{maximum}{
    type=subscript,
    name={\ensuremath{max}},
    description={Indicates the maximum value of a sample}
}
\newglossaryentry{minimum}{
    type=subscript,
    name={\ensuremath{min}},
    description={Indicates the minimum value of a sample}
}
\newglossaryentry{percentil95}{
    type=subscript,
    name={\ensuremath{95}},
    description={Indicates the 95-percentile of a distribution}
}
\newglossaryentry{percentil99}{
    type=subscript,
    name={\ensuremath{99}},
    description={Indicates the 99 percentil of a distribution}
}
\newglossaryentry{solid}{
    type=superscript,
    name={\ensuremath{\mathrm{s}}},
    description={Used for solid related variables}
}
\newglossaryentry{fluid}{
    type=superscript,
    name={\ensuremath{\mathrm{f}}},
    description={Used for fluid related variables}
}
\newglossaryentry{fs}{
    type=superscript,
    name={\ensuremath{\mathrm{fs}}},
    description={Used for variables related to the fluid-solid interface}
}
\newglossaryentry{zeroTime}{
    type=superscript,
    name={\ensuremath{\mathrm{0}}},
    description={Indicates variables at the initial instant $t = 0$}
}
\newglossaryentry{timeIndex}{
    type=superscript,
    name={\ensuremath{\mathrm{n}}},
    description={Time index}
}
\newglossaryentry{fsiIteration}{
    type=subscript,
    name={\ensuremath{k}},
    description={%
        Index for a coupling iteration of the outer loop of the IQN-ILS
        algorithm
    }
}
\newglossaryentry{iteration}{
    type=hidden,
    name={\ensuremath{m}},
    description={%
        Index for iterations when solving a linear system  $\matrix{A}
        \cvec{x} = \cvec{b}$ with an iterative method
    }
}
\newglossaryentry{normal}{
    type=subscript,
    name={\ensuremath{n}},
    description={%
        Indicate the normal component of a vector to a surface, i.e.
        $\vec{u}_{n} = \vec{n}(\vec{n} \dprod \vec{u})$
    }
}
\newglossaryentry{tangential}{
    type=subscript,
    name={\ensuremath{t}},
    description={%
        Indicate the tangential component of a vector to a surface, i.e.
        $\vec{u}_{t} = (\tensor{I} - \vec{n}\vec{n}) \dprod \vec{u}$
    }
}
\newglossaryentry{boundary}{
    type=subscript,
    name={\ensuremath{b}},
    description={%
        Indicate a property on a boundary of a discretized domain in the
        context of the Finite Volume Method.
    }
}

\newglossaryentry{time}{
    type=latin,
    name={\ensuremath{t}},
    description={Time},
    unit={\si{\second}}
}
\newglossaryentry{temperature}{
    type=latin,
    name={\ensuremath{T}},
    description={Temperature},
    unit={\si{\kelvin}}
}
\newglossaryentry{deltat}{
    type=greek,
    name={\ensuremath{\Delta \gls{time}}},
    description={Time-step},
    unit={\si{\second}}
}
\newglossaryentry{mass}{
    type=latin,
    name={\ensuremath{m}},
    description={Mass},
    unit={\si{\kilo\gram}}
}
\newglossaryentry{length}{
    type=latin,
    name={\ensuremath{L}},
    description={Length},
    unit={\si{\meter}}
}
\newglossaryentry{perimeter}{
    type=latin,
    name={\ensuremath{P}},
    description={Perimeter of a contour},
    unit={\si{\meter}}
}
\newglossaryentry{surfaceArea}{
    type=latin,
    name={\ensuremath{A}},
    description={Area of a surface or region},
    unit={\si{\square\meter}}
}
\newglossaryentry{volume}{
    type=latin,
    name={\ensuremath{V}},
    description={Volume of a region in spatial coordinates},
    unit={\si{\cubic\meter}}
}
\newglossaryentry{surface}{
    type=latin,
    name={\ensuremath{S}},
    description={%
        Surface of volume \gls{volume}, $\partial V$, in spatial coordinates
    },
    unit={\si{\square\meter}}
}
\newglossaryentry{density}{
    type=greek,
    name={\ensuremath{\rho}},
    description={Density},
    unit={\si{\kilo\gram\per\cubic\meter}}
}
\newglossaryentry{force}{
    type=latin,
    name={\ensuremath{\vec{f}}},
    description={Force vector},
    unit={\si{\newton}}
}
\newglossaryentry{linearMomentum}{
    type=latin,
    name={\ensuremath{\vec{p}}},
    description={Linear momentum vector},
    unit={\si{\kilo\gram\meter\per\second}}
}
\newglossaryentry{forceComp}{
    type=hidden,
    name={\ensuremath{f}},
    description={Component of the force vector},
    unit={\si{\newton}}
}
\newglossaryentry{traction}{
    type=latin,
    name={\ensuremath{\vec{t}}},
    description={Traction vector},
    unit={\si{\newton\per\square\meter}}
}
\newglossaryentry{CauchyStress}{
    type=greek,
    name={\ensuremath{\tensor{\sigma}}},
    description={Cauchy stress tensor},
    unit={\si{\pascal}}
}
\newglossaryentry{CauchyStressComp}{
    type=hidden,
    name={\ensuremath{\sigma}},
    description={Cauchy stress tensor component},
    unit={\si{\pascal}}
}
\newglossaryentry{bodyForcesPerVolume}{
    type=latin,
    name={\ensuremath{\vec{f}_b}},
    description={Body forces per unit volume},
    unit={\si{\newton\per\cubic\meter}}
}
\newglossaryentry{bodyForcesPerMass}{
    type=latin,
    name={\ensuremath{\vec{b}}},
    description={Body forces per unit mass},
    unit={\si{\newton\per\kilo\gram}}
}
\newglossaryentry{bodyForcesPerMassComp}{
    type=hidden,
    name={\ensuremath{{b}}},
    description={Body forces per unit mass component},
    unit={\si{\newton\per\kilo\gram}}
}
\newglossaryentry{gravity}{
    type=latin,
    name={\ensuremath{\vec{g}}},
    description={Gravitational acceleration vector},
    unit={\si{\meter\per\square\second}}
}
\newglossaryentry{gravityComp}{
    type=hidden,
    name={\ensuremath{g}},
    description={Gravitational acceleration vector},
    unit={\si{\meter\per\square\second}}
}

\newglossaryentry{particle}{
    type=latin,
    name={\ensuremath{\xi}},
    description={Representation of a material particle},
    unit={-}
}
\newglossaryentry{EulerCoord}{
    type=latin,
    name={\ensuremath{\vec{x}}},
    description={Eulerian or spatial coordinates system (Cartesian coordinates)},
    unit={\si{\meter}}
}
\newglossaryentry{meshCoord}{
    type=greek,
    name={\ensuremath{\gvec{\chi}}},
    description={%
        Referential coordinates system for the arbitrary Lagrangian-Eulerian
        description
    },
    unit={\si{\meter}}
}
\newglossaryentry{LagrangeCoord}{
    type=greek,
    name={\ensuremath{\gvec{\xi}}},
    description={Lagrangian or material coordinates system},
    unit={\si{\meter}}
}
\newglossaryentry{meshCoordComp}{
    type=hidden,
    name={\ensuremath{\chi}},
    description={Referential, ALE or mesh coordinates system components},
    unit={\si{\meter}}
}
\newglossaryentry{EulerCoordComp}{
    type=hidden,
    name={\ensuremath{x}},
    description={Eulerian or spatial coordinates components},
    unit={\si{\meter}}
}
\newglossaryentry{LagrangeCoordComp}{
    type=hidden,
    name={\ensuremath{\xi}},
    description={Lagrangian or material coordinates components},
    unit={\si{\meter}}
}

\newglossaryentry{domain}{
    type=greek,
    name={\ensuremath{\Omega}},
    description={Spatial solution domain},
    unit={\si{\cubic\meter}}
}
\newglossaryentry{domainBoundary}{
    type=greek,
    name={\ensuremath{\Gamma}},
    description={Boundary of the domain \gls{domain}},
    unit={\si{\square\meter}}
}
\newglossaryentry{contour}{
    type=greek,
    name={\ensuremath{\Gamma}},
    description={%
        Generic three-dimensional contour; boundary of an open surface
        \gls{surface}, $\partial \gls{surface}$, in spatial coordinates
    },
    unit={\si{\meter}}
}
\newglossaryentry{nablaSpatial}{
    type=other,
    name={\ensuremath{\nabla}},
    description={Nabla operator in Eulerian coordinates},
    unit={\si{\per\meter}}
}
\newglossaryentry{nablaMaterial}{
    type=other,
    name={\ensuremath{\nabla_0}},
    description={Nabla operator in the initial material coordinates},
    unit={\si{\per\meter}}
}
\newglossaryentry{materialDomain}{
    type=greek,
    name={\ensuremath{\Omega_{\xi}}},
    description={Solution domain as seen by the material coordinates},
    unit={\si{\cubic\meter}}
}
\newglossaryentry{meshDomain}{
    type=greek,
    name={\ensuremath{\Omega_{\chi}}},
    description={Solution domain as seen by the reference coordinates},
    unit={\si{\cubic\meter}}
}
\newglossaryentry{materialVolume}{
    type=latin,
    name={\ensuremath{V_{\xi}}},
    description={Volume of an amount of particles in material coordinates},
    unit={\si{\cubic\meter}}
}
\newglossaryentry{meshVolume}{
    type=latin,
    name={\ensuremath{V_{\chi}}},
    description={Control volume in ALE coordinates},
    unit={\si{\cubic\meter}}
}
\newglossaryentry{sNormalVector}{
    type=latin,
    name={\ensuremath{\vec{n}}},
    description={%
        Unit normal vector to the surface \gls{surface} pointing outwards
    },
    unit={-}
}

\newglossaryentry{funcMaterial}{
    type=latin,
    name={\ensuremath{f_{\xi}}},
    description={Function $f$ representing any scalar variable of the material},
    unit={-}
}
\newglossaryentry{funcSpatial}{
    type=latin,
    name={\ensuremath{f}},
    description={%
        Spatial function $f$ representing any scalar variable of the material
    },
    unit={-}
}
\newglossaryentry{funcMesh}{
    type=latin,
    name={\ensuremath{f_{\chi}}},
    description={%
        ALE description of a function $f$ representing any scalar variable of
        the material
    },
    unit={-}
}
\newglossaryentry{materialSpatialMap}{
    type=greek,
    name={\ensuremath{\gvec{\Phi}}},
    description={Mapping of material coordinates onto spatial coordinates},
    unit={-}
}
\newglossaryentry{meshSpatialMap}{
    type=greek,
    name={\ensuremath{\gvec{\Psi}}},
    description={Mapping of mesh coordinates onto spatial coordinates},
    unit={-}
}
\newglossaryentry{materialMeshMap}{
    type=greek,
    name={\ensuremath{\gvec{\Xi}}},
    description={Mapping of material coordinates onto mesh coordinates},
    unit={-}
}
\newglossaryentry{identityTensor}{
    type=latin,
    name={\ensuremath{\tensor{I}}},
    description={Second order identity tensor},
    unit={-}
}
\newglossaryentry{kroneckerDelta}{
    type=hidden,
    name={\ensuremath{\delta_{ij}}},
    description={Kronecker's delta},
    unit={-}
}
\newglossaryentry{fourthIdentityTensor}{
    type=latin,
    name={\ensuremath{\fourthTensor{I}}},
    description={Fourth-order identity tensor},
    unit={-}
}

\newglossaryentry{Jacobian}{
    type=latin,
    name={\ensuremath{J}},
    description={Jacobian determinant of a mapping},
    unit={-}
}
\newglossaryentry{materialJacobianMatrix}{
    type=latin,
    name={\ensuremath{\mathbf{J}_{\gls{materialSpatialMap}}}},
    description={Jacobian matrix of the material-spatial mapping},
    unit={-}
}
\newglossaryentry{meshJacobianMatrix}{
    type=latin,
    name={\ensuremath{\mathbf{J}_{\gls{meshSpatialMap}}}},
    description={Jacobian matrix of the mesh-spatial mapping},
    unit={-}
}
\newglossaryentry{materialJacobian}{
    type=latin,
    name={\ensuremath{\mathrm{J}_{\gls{materialSpatialMap}}}},
    description={Jacobian of the material-spatial mapping},
    unit={-}
}
\newglossaryentry{meshJacobian}{
    type=latin,
    name={\ensuremath{\mathrm{J}_{\gls{meshSpatialMap}}}},
    description={Jacobian of the mesh-spatial mapping},
    unit={-}
}

\newglossaryentry{velocity}{
    type=latin,
    name={\ensuremath{\vec{v}}},
    description={Material velocity},
    unit={\si{\meter\per\second}}
}
\newglossaryentry{angularVelocity}{
    type=latin,
    name={\ensuremath{\omega}},
    description={Angular velocity},
    unit={\si{\radian\per\second}}
}
\newglossaryentry{velocityComp}{
    type=hidden,
    name={\ensuremath{v}},
    description={Material velocity},
    unit={\si{\meter\per\second}}
}
\newglossaryentry{xVelocity}{
    type=latin,
    name={\ensuremath{v_x}},
    description={%
        Velocity component on the x direction of the spatial coordinates
    },
    unit={\si{\meter\per\second}}
}
\newglossaryentry{yVelocity}{
    type=latin,
    name={\ensuremath{v_y}},
    description={%
        Velocity component on the y direction of the spatial coordinates
    },
    unit={\si{\meter\per\second}}
}
\newglossaryentry{zVelocity}{
    type=latin,
    name={\ensuremath{v_z}},
    description={%
        Velocity component on the z direction of the spatial coordinates
    },
    unit={\si{\meter\per\second}}
}
\newglossaryentry{meshVelocity}{
    type=greek,
    name={\ensuremath{\gvec{\omega}}},
    description={Mesh velocity of a cell centroid},
    unit={\si{\meter\per\second}}
}
\newglossaryentry{xMeshVelocity}{
    type=hidden,
    name={\ensuremath{\omega_x}},
    description={%
        Mesh velocity component on the x direction of the spatial coordinates
    },
    unit={\si{\meter\per\second}}
}
\newglossaryentry{yMeshVelocity}{
    type=hidden,
    name={\ensuremath{\omega_y}},
    description={%
        Mesh velocity component on the y direction of the spatial coordinates
    },
    unit={\si{\meter\per\second}}
}
\newglossaryentry{zMeshVelocity}{
    type=hidden,
    name={\ensuremath{\omega_z}},
    description={%
        Mesh velocity component on the z direction of the spatial coordinates
    },
    unit={\si{\meter\per\second}}
}
\newglossaryentry{meshVelocityNode}{
    type=greek,
    name={\ensuremath{\gvec{\omega}_{node}}},
    description={%
        Mesh velocity at the geometric nodes of the Finite Volume mesh
    },
    unit={\si{\meter\per\second}}
}
\newglossaryentry{materialMeshVelocity}{
    type=latin,
    name={\ensuremath{\hat{\vec{v}}}},
    description={%
        Material velocity measured from the ALE coordinates point of view
    },
    unit={\si{\meter\per\second}}
}
\newglossaryentry{convectiveVelocity}{
    type=latin,
    name={\ensuremath{\vec{c}}},
    description={Convective velocity},
    unit={\si{\meter\per\second}}
}

\newglossaryentry{massFlowRate}{
    type=latin,
    name={\ensuremath{\dot{m}}},
    description={Mass flow rate},
    unit={\si{\kilo\gram\per\second}}
}

\newglossaryentry{numberMoles}{
    type=latin,
    name={\ensuremath{N}},
    description={Number of moles of a species $i$ in a system},
    unit={\si{\mol}}
}
\newglossaryentry{numberSpecies}{
    type=latin,
    name={\ensuremath{n}},
    description={Number of moles of a species $i$ in a system},
    unit={}
}

\newglossaryentry{specificVolume}{
    type=latin,
    name={\ensuremath{v}},
    description={Specific volume -- volume per mass},
    unit={\si{\cubic\meter\per\kilogram}}
}

\newglossaryentry{internalEnergy}{
    type=latin,
    name={\ensuremath{U}},
    description={Internal energy of a system},
    unit={\si{\joule}}
}
\newglossaryentry{intInternalEnergy}{
    type=latin,
    name={\ensuremath{u}},
    description={Intensive internal energy of a system},
    unit={\si{\joule\per\kilo\gram}}
}

\newglossaryentry{enthalpy}{
    type=latin,
    name={\ensuremath{H}},
    description={Enthalpy of a system},
    unit={\si{\joule}}
}
\newglossaryentry{intEnthalpy}{
    type=latin,
    name={\ensuremath{h}},
    description={Intensive enthalpy of a system},
    unit={\si{\joule\per\kilo\gram}}
}

\newglossaryentry{entropy}{
    type=latin,
    name={\ensuremath{S}},
    description={Entropy of a system},
    unit={\si{\joule\per\kelvin}}
}
\newglossaryentry{intEntropy}{
    type=latin,
    name={\ensuremath{s}},
    description={Intensive entropy of a system},
    unit={\si{\joule\per\kilo\gram\per\kelvin}}
}

\newglossaryentry{HelmholtzFreeEnergy}{
    type=latin,
    name={\ensuremath{F}},
    description={Helmholtz free-energy of a system},
    unit={\si{\joule}}
}
\newglossaryentry{intHelmholtzFreeEnergy}{
    type=latin,
    name={\ensuremath{f}},
    description={Intensive Helmholtz free-energy of a system},
    unit={\si{\joule\per\kilo\gram}}
}

\newglossaryentry{GibbsFreeEnergy}{
    type=latin,
    name={\ensuremath{G}},
    description={Gibbs free-energy of a system},
    unit={\si{\joule}}
}
\newglossaryentry{intGibbsFreeEnergy}{
    type=latin,
    name={\ensuremath{g}},
    description={Intensive Gibbs free-energy of a system},
    unit={\si{\joule\per\kilo\gram}}
}

\newglossaryentry{entropyCreated}{
    type=latin,
    name={\ensuremath{S_{ger}}},
    description={%
        Entropy created or generated in a process undergone by a system
    },
    unit={\si{\joule\per\kelvin}}
}
\newglossaryentry{intEntropyCreated}{
    type=latin,
    name={\ensuremath{s_{ger}}},
    description={%
        Intensive entropy created or generated in a process undergone by a
        system
    },
    unit={\si{\joule\per\kelvin}}
}
\newglossaryentry{entropyCreatedRate}{
    type=latin,
    name={\ensuremath{\dot{S}_{ger}}},
    description={%
        Entropy created or generated in a process undergone by a system
    },
    unit={\si{\watt\per\kelvin}}
}
\newglossaryentry{intEntropyCreatedRate}{
    type=latin,
    name={\ensuremath{\dot{s}_{ger}}},
    description={%
        Intensive entropy created or generated in a process undergone by a
        system
    },
    unit={\si{\watt\per\kelvin}}
}

\newglossaryentry{chemicalPotential}{
    type=greek,
    name={\ensuremath{\mu}},
    description={%
        Chemical potential of a species crossing the boundary of a system
    },
    unit={\si{\joule\per\kelvin}}
}
\newglossaryentry{intIrreversibility}{
    type=latin,
    name={\ensuremath{I}},
    description={Intensive irreversibility irreversibility},
    unit={\si{\joule}}
}
\newglossaryentry{thermodynamicSubstate}{
    type=greek,
    name={\ensuremath{\upsilon}},
    description={%
        Generic thermodynamic sub-state of a system, identified as the
        intensive counterpart of the generalized displacement producing work
        transfers
    },
    unit={-}
}
\newglossaryentry{generalizedForce}{
    type=greek,
    name={\ensuremath{\tau}},
    description={Generalized force applied to a system},
    unit={-}
}
\newglossaryentry{generalizedDisplacement}{
    type=greek,
    name={\ensuremath{\Upsilon}},
    description={Generalized displacement applied to a system},
    unit={-}
}

\newglossaryentry{heatTransfer}{
    type=latin,
    name={\ensuremath{Q}},
    description={Heat transfer interaction on the boundary of a system},
    unit={\si{\joule}}
}
\newglossaryentry{heatFlowVector}{
    type=latin,
    name={\ensuremath{\vec{q}}},
    description={%
        Heat flux or heat flow vector, defined as the heat transfer per unit
        surface area
    },
    unit={\si{\watt\per\square\meter}}
}

\newglossaryentry{workTransfer}{
    type=latin,
    name={\ensuremath{W}},
    description={Generic work transfer interaction on the boundary of a system},
    unit={\si{\joule}}
}
\newglossaryentry{internalMechanicalPower}{
    type=latin,
    name={\ensuremath{\dot{w}_{int}}},
    description={%
        Internal mechanical power in a system, caused by the stress and
        deformation
    },
    unit={\si{\watt}}
}

\newglossaryentry{internalMechanicalWork}{
    type=latin,
    name={\ensuremath{w_{int}}},
    description={%
        Internal mechanical work in a system, caused by the stress and
        deformation
    },
    unit={\si{\joule}}
}
\newglossaryentry{numberWorkSources}{
    type=latin,
    name={\ensuremath{m}},
    description={Number of work sources to a system},
    unit={}
}

\newglossaryentry{isothermalCompressibility}{
    type=greek,
    name={\ensuremath{\psi}},
    description={Isothermal compressibility of a fluid},
    unit={\si{\square\second\per\square\meter}}
}

\newglossaryentry{bulkModulus}{
    type=greek,
    name={\ensuremath{\kappa}},
    description={Bulk modulus},
    unit={\si{\pascal}}
}

\newglossaryentry{cvNumber}{
    type=latin,
    name={\ensuremath{N_{CV}}},
    description={Number of control volumes in a finite volume mesh},
    unit={-}
}

\newglossaryentry{centroid}{
    type=subscript,
    name={\ensuremath{C}},
    description={%
        Indicates variables evaluated at the centroid of a control volume
    }
}
\newglossaryentry{nbCentroid}{
    type=subscript,
    name={\ensuremath{N}},
    description={%
        Indicates variables evaluated at the centroid of a neighbor control
        volume to cell $V_C$
    }
}
\newglossaryentry{owner}{
    type=subscript,
    name={\ensuremath{\gls{centroid}}},
    description={%
        Indicates variables evaluated at the centroid of a the face owner
        control volume
    }
}
\newglossaryentry{neighbor}{
    type=subscript,
    name={\ensuremath{\gls{nbCentroid}}},
    description={%
        Indicates variables evaluated at the centroid of a face neighbor
        control volume
    }
}
\newglossaryentry{face}{
    type=subscript,
    name={\ensuremath{f}},
    description={%
        Indicate a face of a polyhedral control volume and a property evaluated
        at it
    }
}
\newglossaryentry{skewFace}{
    type=subscript,
    name={\ensuremath{\gls{face}^{\prime}}},
    description={%
        Indicates the point where the vector joining two face adjacent cell's
        centroids intercepts the face
    }
}

\newglossaryentry{fvCellVolume}{
    type=latin,
    name={\ensuremath{\gls{volume}_{\gls{centroid}}}},
    description={Volume of a cell with centroid $C$ in a finite volume mesh},
    unit={\si{\cubic\meter}}
}
\newglossaryentry{fvCellSurface}{
    type=latin,
    name={\ensuremath{\gls{surface}_{\gls{centroid}}}},
    description={Surface of a cell with centroid $C$ in a finite volume mesh},
    unit={\si{\square\meter}}
}
\newglossaryentry{fvFaceNormal}{
    type=latin,
    name={\ensuremath{\vec{S}_{\gls{face}}}},
    description={Normal vector to the face $f$ of a polyhedral cell},
    unit={\si{\square\meter}}
}
\newglossaryentry{joinCentroidVector}{
    type=latin,
    name={\ensuremath{\vec{d}_{\gls{centroid}\gls{nbCentroid}}}},
    description={%
        Vector joining the centroids of adjacent cells in a finite volume mesh
    },
    unit={\si{\meter}}
}
\newglossaryentry{skewnessVector}{
    type=latin,
    name={\ensuremath{\vec{m}}},
    description={%
        Vector joining the face centroid with the interception point between
        the face and the line between the owner and neighbour cells.
    },
    unit={\si{\meter}}
}
\newglossaryentry{facePyramidHeigth}{
    type=latin,
    name={\ensuremath{\vec{h}_{\gls{face}\gls{centroid}}}},
    description={%
        Heigth of a cell face pyramid -- is the opposite of the vector joining
        the cell centroid and the face centroid
    },
    unit={\si{\meter}}
}
\newglossaryentry{centroidFaceVector}{
    type=latin,
    name={\ensuremath{\vec{d}_{\gls{centroid}\gls{face}}}},
    description={%
        Vector joining the centroids of a cell \gls{fvCellVolume} and the
        centroid of a face \gls{face}
    },
    unit={\si{\meter}}
}
\newglossaryentry{ownerFaceVector}{
    type=latin,
    name={\ensuremath{\vec{d}_{\gls{owner}\gls{face}}}},
    description={%
        Vector joining the centroid of a face owner cell and the centroid of
        the face \gls{face}
    },
    unit={\si{\meter}}
}
\newglossaryentry{neighFaceVector}{
    type=latin,
    name={\ensuremath{\vec{d}_{\gls{neighbor}\gls{face}}}},
    description={%
        Vector joining the centroid of a face neighbor cell and the centroid of
        the face \gls{face}
    },
    unit={\si{\meter}}
}

\newglossaryentry{projCentroidNormalVec}{
    type=hidden,
    name={\ensuremath{\vec{d}_{fn}}},
    description={%
        Vector formed by the projection of \gls{joinCentroidVector}
        on \gls{sNormalVector}
    },
    unit={\si{\meter}}
}
\newglossaryentry{projCentroidNormalNorm}{
    type=latin,
    name={\ensuremath{d_{fn}}},
    description={%
        Projection of the vector \gls{joinCentroidVector} on the face normal
        \gls{sNormalVector}
    },
    unit={\si{\meter}}
}
\newglossaryentry{advWeightFactor}{
    type=hidden,
    name={\ensuremath{w}},
    description={%
        Weight of cell \gls{centroid} on the advective interpolation profile
    },
    unit={-}
}
\newglossaryentry{faceVolumeSwept}{
    type=latin,
    name={\ensuremath{\dot{V}_f}},
    description={%
        Volume swept by face \gls{face} of control volume \gls{fvCellVolume}
    },
    unit={\si{\cubic\meter\per\second}}
}
\newglossaryentry{normalDecomposition1}{
    type=latin,
    name={\ensuremath{\vec{\beta}}},
    description={%
        Vector joining the centroid of face \gls{face} and the neighbor cell
        centroid \gls{nbCentroid}
    },
    unit={-}
}
\newglossaryentry{normalDecomposition1Norm}{
    type=hidden,
    name={\ensuremath{\beta}},
    description={%
        Modulus of the vector joining the centroid of face \gls{face} and the
        neighbor cell centroid \gls{nbCentroid}
    },
    unit={-}
}
\newglossaryentry{normalDecomposition2}{
    type=greek,
    name={\ensuremath{\vec{\kappa}}},
        description={Complement vector of the decomposition of
        \gls{fvFaceNormal} in \gls{normalDecomposition1}},
        unit={-}
}

\newglossaryentry{transportVariable}{
    type=greek,
    name={\ensuremath{\phi}},
    description={%
        Generic intensive (per unit mass) variable transported by a fluid flow
    },
    unit={$[\phi]/$\si{\kilo\gram}}
}
\newglossaryentry{integralTransportVariable}{
    type=greek,
    name={\ensuremath{\Phi}},
    description={%
        Generic intensive variable transported by a fluid flow integrated in a
        finite control volume
    },
    unit={$[\Phi]$}
}
\newglossaryentry{transportVariablePerVolume}{
    type=greek,
    name={\ensuremath{\varphi}},
    description={%
        Generic intensive (per unit volume) variable transported by a fluid
        flow
    },
    unit={$[\Phi]/$\si{\cubic\meter}}
}
\newglossaryentry{transportDiffusivity}{
    type=greek,
    name={\ensuremath{\Gamma^{\phi}}},
    description={Diffusivity of \gls{transportVariable}},
    unit={\si{\square\meter\per\second}}
}
\newglossaryentry{transportSource}{
    type=latin,
    name={\ensuremath{S^{\phi}}},
    description={%
        Source and/or sink terms of the differential transport equation of
        \gls{transportVariable}
    },
    unit={$[\phi]/$\si{\cubic\meter\per\second}}
}
\newglossaryentry{linearTransportSource}{
    type=latin,
    name={\ensuremath{\bar{S}^{\phi}}},
    description={%
        Mean source and/or sink terms of the integral transport equation of
        \gls{transportVariable}
    },
    unit={$[\phi]/$\si{\cubic\meter\per\second}}
}
\newglossaryentry{fvCoeffs}{
    type=latin,
    name={\ensuremath{a}},
    description={%
        Generic coefficients of the linearized transport equation for the
        volume $V_C$
    },
    unit={\si{\kilo\gram\per\second}}
}
\newglossaryentry{fvDiagCoeffs}{
    type=latin,
    name={\ensuremath{\gls{fvCoeffs}_{\gls{centroid}}}},
    description={%
        Diagonal coefficients of the linearized transport equation for the
        volume $V_C$
    },
    unit={\si{\kilo\gram\per\second}}
}
\newglossaryentry{fvOffDiagCoeffs}{
    type=latin,
    name={\ensuremath{\gls{fvCoeffs}_{\gls{nbCentroid}}}},
    description={%
        Off-diagonal coefficients of the linearized transport equation for the
        volume $V_C$
    },
    unit={\si{\kilo\gram\per\second}}
}
\newglossaryentry{fvMatrix}{
    type=latin,
    name={\ensuremath{\matrix{A}}},
    description={%
        Coefficients matrix of a system of equations assembled for a finite
        volume mesh
    },
    unit={$[\phi]$}
}
\newglossaryentry{sourceScalar}{
    type=latin,
    name={\ensuremath{b}},
    description={%
        Source term of the linearized transport equation for a control volume
        $V_C$
    },
    unit={$[\phi]$\si{\kilo\gram\per\second}}
}
\newglossaryentry{fvSource}{
    type=hidden,
    name={\ensuremath{\cvec{b}}},
    description={%
        Source term of the linearized system equation for a finite volume mesh
    },
    unit={-}
}
\newglossaryentry{sourceVector}{
    type=latin,
    name={\ensuremath{\vec{b}}},
    description={%
        Source vector of the linearized momentum equations for a control volume
    },
    unit={$[\phi]$\si{\kilo\gram\per\second}}
}
\newglossaryentry{pressureCorrection}{
    type=latin,
    name={\ensuremath{p'}},
    description={%
        Pressure correction field in the pressure-velocity algorithms to solve
        the coupled Navier-Stokes equations
    },
    unit={\si{\pascal}}
}
\newglossaryentry{transportOperator}{
    type=hidden,
    name={\ensuremath{\matrix{H}}},
    description={Transport operator},
    unit={-}
}
\newglossaryentry{sourceOperator}{
    type=hidden,
    name={\ensuremath{\matrix{B}}},
    description={Source operator},
    unit={-}
}
\newglossaryentry{pressureDiffOperator}{
    type=hidden,
    name={\ensuremath{\matrix{D}}},
    description={Pressure diffusive operator},
    unit={-}
}

\newglossaryentry{meshNonOrthog}{
    type=latin,
    name={\ensuremath{mesh_{nonOrtho}}},
    description={Mesh non-orthogonality}
}

\newglossaryentry{meshSkewness}{
    type=latin,
    name={\ensuremath{mesh_{skew}}},
    description={Mesh skewness}
}

\newglossaryentry{faceTwist}{
    type=latin,
    name={\ensuremath{mesh_{faceTwist}}},
    description={Mesh face twist}
}

\newglossaryentry{faceWeight}{
    type=latin,
    name={\ensuremath{mesh_{faceWeight}}},
    description={Mesh face weight}
}

\newglossaryentry{faceVolumeRatio}{
    type=latin,
    name={\ensuremath{mesh_{faceVolRatio}}},
    description={Mesh face volume ratio}
}

\newglossaryentry{faceTriangTwist}{
    type=latin,
    name={\ensuremath{mesh_{faceTriTwist}}},
    description={Mesh face twist}
}

\newglossaryentry{cellConvexity}{
    type=latin,
    name={\ensuremath{mesh_{cellConcave}}},
    description={Mesh cells convexity: it measures if a cell is concave.}
}

\newglossaryentry{pyramidVolume}{
    type=latin,
    name={\ensuremath{mesh_{cellPyrVol}}},
    description={Mesh cells pyramids volume}
}

\newglossaryentry{cellTetQuality}{
    type=latin,
    name={\ensuremath{mesh_{tetQuality}}},
    description={Mesh cells tetrahedral decomposition quality}
}
\newglossaryentry{polyCellTensor}{
    type=latin,
    name={\ensuremath{\tensor{T}}},
    description={%
        Second-order tensor formed by the outer product of the face vector of a
        generic polyhedral cell
    }
}
\newglossaryentry{cellDeterminant}{
    type=latin,
    name={\ensuremath{mesh_{cellDeterminant}}},
    description={Mesh face twist}
}

\newglossaryentry{dynamicViscosity}{
    type=greek,
    name={\ensuremath{\mu^{\gls{fluid}}}},
    description={Dynamic viscosity of a fluid},
    unit={\si{\pascal\second}}
}
\newglossaryentry{secondViscosity}{
    type=greek,
    name={\ensuremath{\lambda^{\gls{fluid}}}},
    description={%
        Second viscosity of a fluid (sometimes referred to as
        \emph{bulk viscosity})
    },
    unit={\si{\pascal\second}}
}
\newglossaryentry{kinematicViscosity}{
    type=greek,
    name={\ensuremath{\nu^{\gls{fluid}}}},
    description={Kinematic viscosity of a fluid},
    unit={\si{\square\meter\per\second}}
}
\newglossaryentry{pressure}{
    type=latin,
    name={\ensuremath{p}},
    description={Pressure field of a fluid flow},
    unit={\si{\pascal}}
}
\newglossaryentry{meshDiffusivity}{
    type=greek,
    name={\ensuremath{\gamma}},
    description={%
        Mesh diffusivity for the Laplace equation governing the mesh motion
    },
    unit={-}
}
\newglossaryentry{centroidToBoundaryDistance}{
    type=latin,
    name={\ensuremath{l}},
    description={Cell centroid distance to a selected boundary},
    unit={\si{\meter}}
}
\newglossaryentry{viscousCauchyStress}{
    type=greek,
    name={\ensuremath{\gvec{\tau}}},
    description={Viscous part of the Cauchy stress tensor of a fluid flow},
    unit={\si{\pascal}}
}
\newglossaryentry{viscousCauchyStressComp}{
    type=hidden,
    name={\ensuremath{\tau}},
    description={Viscous part of the Cauchy stress tensor of a fluid flow},
    unit={\si{\pascal}}
}
\newglossaryentry{ReynoldsNumber}{
    type=latin,
    name={\ensuremath{Re}},
    description={Reynolds number},
    unit={-}
}
\newglossaryentry{CourantNumber}{
    type=latin,
    name={\ensuremath{Co}},
    description={Courant number},
    unit={-}
}
\newglossaryentry{shearRate}{
    type=latin,
    name={\ensuremath{\dot{\gamma}}},
    description={%
        Shear rate component in the flow between two coaxial cylinders
    },
    unit={\si{\per\second}}
}
\newglossaryentry{yieldStress}{
    type=latin,
    name={\ensuremath{\tau_C}},
    description={Yield stress of a Herschel-Bulkley fluid},
    unit={\si{\pascal}}
}
\newglossaryentry{consistency}{
    type=latin,
    name={\ensuremath{k_n}},
    description={Consistency of a Herschel-Bulkley fluid},
    unit={\si{\pascal\second^n}}
}
\newglossaryentry{indexFlow}{
    type=latin,
    name={\ensuremath{n}},
    description={Index flow of a Herschel-Bulkley fluid},
    unit={-}
}
\newglossaryentry{apparentViscosity}{
    type=greek,
    name={\ensuremath{\eta^{\gls{fluid}}}},
    description={Apparent viscosity of a non-Newtonian fluid},
    unit={\si{\pascal\second}}
}
\newglossaryentry{CassonCoefficient}{
    type=latin,
    name={\ensuremath{k}},
    description={Consistency coefficient of Casson's model},
    unit={\si{\pascal\second}}
}

\newglossaryentry{firstInvariant}{
    type=latin,
    name={\ensuremath{I_1}},
    description={First principal invariant of a second order tensor},
    unit={-}
}
\newglossaryentry{secondInvariant}{
    type=latin,
    name={\ensuremath{I_2}},
    description={Second principal invariant of a second-order tensor},
    unit={-}
}
\newglossaryentry{thirdInvariant}{
    type=latin,
    name={\ensuremath{I_3}},
    description={Third principal invariant of a second order tensor},
    unit={-}
}
\newglossaryentry{modifiedSecondInvariant}{
    type=latin,
    name={\ensuremath{J_2}},
    description={Modified second invariant of the rate of deformation tensor},
    unit={-}
}
\newglossaryentry{elasticModuliTensor}{
    type=latin,
    name={\ensuremath{\fourthTensor{C}}},
    description={Elastic moduli tensor},
    unit={\si{\pascal}}
}
\newglossaryentry{elasticModuliComp}{
    type=latin,
    name={\ensuremath{C}},
    description={Elastic moduli tensor component},
    unit={\si{\pascal}}
}
\newglossaryentry{YoungModulus}{
    type=latin,
    name={\ensuremath{E}},
    description={Young's modulus of a solid},
    unit={\si{\pascal}}
}
\newglossaryentry{shearModulus}{
    type=latin,
    name={\ensuremath{G}},
    description={Shear modulus of a solid},
    unit={\si{\pascal}}
}
\newglossaryentry{linearShearModulus}{
    type=latin,
    name={\ensuremath{\mu}},
    description={Shear modulus of the linearized theory of Solid Mechanics},
    unit={\si{\pascal}}
}
\newglossaryentry{PoissonRatio}{
    type=greek,
    name={\ensuremath{\nu^{\gls{solid}}}},
    description={Poisson's ratio of a solid},
    unit={-}
}
\newglossaryentry{strainEnergyFunction}{
    type=greek,
    name={\ensuremath{\Psi}},
    description={%
        Deformation work function or strain energy function -- function of
        \gls{GLstrain}. Energy from deformation work per unit volume
    },
    unit={\si{\joule\per\cubic\meter}}
}
\newglossaryentry{deformationWorkFunction}{
    type=latin,
    name={\ensuremath{\Psi}},
    description={%
        Deformation work function or strain energy function -- function of
        \gls{rightCauchyGreenDeformation}. Energy from deformation work per
        unit volume
    },
    unit={\si{\joule\per\cubic\meter}}
}
\newglossaryentry{lameConst1}{
    type=greek,
    name={\ensuremath{\mu^{\gls{solid}}}},
    description={First Lamé's constant of a solid},
    unit={\si{\pascal}}
}
\newglossaryentry{lameConst2}{
    type=greek,
    name={\ensuremath{\lambda^{\gls{solid}}}},
    description={Second Lamé's constant of a solid},
    unit={\si{\pascal}}
}
\newglossaryentry{nominalStress}{
    type=latin,
    name={\ensuremath{\tensor{P}}},
    description={Nominal stress tensor},
    unit={\si{\pascal}}
}
\newglossaryentry{nominalStressComp}{
    type=hidden,
    name={\ensuremath{P}},
    description={Nominal stress tensor component},
    unit={\si{\pascal}}
}
\newglossaryentry{rateDeformationTensor}{
    type=latin,
    name={\ensuremath{\tensor{D}}},
    description={Rate of deformation tensor},
    unit={\si{\per\second   }}
}
\newglossaryentry{rateDeformationTensorComp}{
    type=hidden,
    name={\ensuremath{D}},
    description={Rate of deformation tensor component},
    unit={\si{\per\second}}
}
\newglossaryentry{deformationGradient}{
    type=latin,
    name={\ensuremath{\tensor{F}}},
    description={Deformation gradient tensor},
    unit={-}
}
\newglossaryentry{rotationTensor}{
    type=latin,
    name={\ensuremath{\tensor{R}}},
    description={Rotation tensor},
    unit={-}
}

\newglossaryentry{stretch}{
    type=greek,
    name={\ensuremath{\lambda}},
    description={%
        Stretch defined as the current length divided by the reference length
        of a solid under deformation
    },
    unit={-}
}
\newglossaryentry{deformationGradComp}{
    type=hidden,
    name={\ensuremath{F}},
    description={Deformation gradient tensor component},
    unit={-}
}
\newglossaryentry{solidDisplacement}{
    type=latin,
    name={\ensuremath{\vec{u}}},
    description={Solid displacement vector},
    unit={m}
}
\newglossaryentry{solidDisplComp}{
    type=hidden,
    name={\ensuremath{u}},
    description={Solid displacement vector component},
    unit={-}
}
\newglossaryentry{xSolidDisplacement}{
    type=hidden,
    name={\ensuremath{u_{\gls{xAxis}}}},
    description={%
        Solid displacement vector component in the x direction of Cartesian
        coordinates
    },
    unit={-}
}
\newglossaryentry{ySolidDisplacement}{
    type=hidden,
    name={\ensuremath{u_{\gls{yAxis}}}},
    description={%
        Solid displacement vector component in the y direction of Cartesian
        coordinates
    },
    unit={-}
}
\newglossaryentry{zSolidDisplacement}{
    type=hidden,
    name={\ensuremath{u_{\gls{zAxis}}}},
    description={%
        Solid displacement vector component in the z direction of Cartesian
        coordinates
    },
    unit={-}
}
\newglossaryentry{2PKstress}{
    type=latin,
    name={\ensuremath{\tensor{S}}},
    description={Second Piola-Kirchhoff stress tensor},
    unit={\si{\pascal}}
}
\newglossaryentry{KirchhoffStress}{
    type=greek,
    name={\ensuremath{\tensor{\sigma}^{\mathrm{K}}}},
    description={Kirchhoff stress tensor},
    unit={\si{\pascal}}
}
\newglossaryentry{KirchhoffStressComp}{
    type=hidden,
    name={\ensuremath{\sigma^{\mathrm{K}}}},
    description={Kirchhoff stress tensor component},
    unit={\si{\pascal}}
}
\newglossaryentry{GLstrain}{
    type=latin,
    name={\ensuremath{\tensor{E}}},
    description={Green-Lagrange strain tensor},
    unit={-}
}
\newglossaryentry{leftCauchyGreenDeformation}{
    type=latin,
    name={\ensuremath{\tensor{B}}},
    description={Left Green-Cauchy deformation tensor},
    unit={-}
}
\newglossaryentry{rightCauchyGreenDeformation}{
    type=latin,
    name={\ensuremath{\tensor{C}}},
    description={Right Green-Cauchy deformation tensor},
    unit={-}
}
\newglossaryentry{2PKstressComp}{
    type=hidden,
    name={\ensuremath{S}},
    description={Second Piola-Kirchhoff stress tensor component},
    unit={\si{\pascal}}
}
\newglossaryentry{GLstrainComp}{
    type=hidden,
    name={\ensuremath{E}},
    description={Green-Lagrange strain tensor component},
    unit={-}
}
\newglossaryentry{leftCauchyGreenDeformationComp}{
    type=hidden,
    name={\ensuremath{B}},
    description={Left Green-Cauchy deformation tensor component},
    unit={-}
}
\newglossaryentry{rightCauchyGreenDeformationComp}{
    type=hidden,
    name={\ensuremath{C}},
    description={Right Green-Cauchy deformation tensor component},
    unit={-}
}
\newglossaryentry{linearStrain}{
    type=greek,
    name={\ensuremath{\tensor{\varepsilon}}},
    description={Linear or engineering strain tensor},
    unit={-}
}
\newglossaryentry{linearStrainComp}{
    type=hidden,
    name={\ensuremath{\varepsilon}},
    description={Linear or engineering strain tensor},
    unit={-}
}

\newglossaryentry{trueLinearStrain}{
    type=greek,
    name={\ensuremath{\tensor{e}}},
    description={True or logarithmic strain tensor},
    unit={-}
}

\newglossaryentry{tensorInvariant}{
    type=hidden,
    name={\ensuremath{I}},
    description={},
    unit={-}
}
\newglossaryentry{fourthInvariant}{
    type=latin,
    name={\ensuremath{I_4}},
    description={%
        Fourth invariant of the right Cauchy-Green deformation tensor with
        direction information
    },
    unit={-}
}
\newglossaryentry{fifthInvariant}{
    type=latin,
    name={\ensuremath{I_5}},
    description={%
        Third invariant of the right Cauchy-Green deformation tensor with
        directional information
    },
    unit={-}
}
\newglossaryentry{sixthInvariant}{
    type=latin,
    name={\ensuremath{I_6}},
    description={%
        Sixth invariant of the right Cauchy-Green deformation tensor with
        directional information
    },
    unit={-}
}
\newglossaryentry{seventhInvariant}{
    type=latin,
    name={\ensuremath{I_7}},
    description={%
        Seventh invariant of the right Cauchy-Green deformation tensor with
        directional information
    },
    unit={-}
}
\newglossaryentry{eighthInvariant}{
    type=latin,
    name={\ensuremath{I_8}},
    description={%
        Eighth invariant of the right Cauchy-Green deformation tensor with
        directional information
    },
    unit={-}
}

\newglossaryentry{MooneyCoeffs}{
    type=hidden,
    name={\ensuremath{c}},
    description={%
        Coefficients of Rivlin series expansion of the strain-energy function
    },
    unit={\si{\kilo\joule\per\cubic\meter}}
}
\newglossaryentry{OgdenCoeffs}{
    type=hidden,
    name={\ensuremath{\alpha}},
    description={Exponents of the Ogden strain-energy function},
    unit={-}
}
\newglossaryentry{FungArgumentFunction}{
    type=latin,
    name={\ensuremath{Q}},
    description={%
        Function of the strain tensor components used in the generic Fung-type
        models
    },
    unit={-}
}
\newglossaryentry{fiberDirection}{
    type=latin,
    name={\ensuremath{\vec{m}}},
    description={%
        Vector of the preferred direction of a fiber reinforced material
    },
    unit={\si{-}}
}
\newglossaryentry{orientationTensor}{
    type=latin,
    name={\ensuremath{\tensor{M}}},
    description={Orientation tensor},
    unit={\si{-}}
}
\newglossaryentry{orientationTensorComp}{
    type=hidden,
    name={\ensuremath{M}},
    description={Orientation tensor component},
    unit={\si{-}}
}
\newglossaryentry{fiberDirectionComp}{
    type=hidden,
    name={\ensuremath{m}},
    description={%
        Vector component of the preferred direction of a fiber reinforced
        material
    },
    unit={\si{-}}
}
\newglossaryentry{groundSubstance}{
    type=subscript,
    name={\ensuremath{g}},
    description={Ground substance}
}
\newglossaryentry{collagen}{
    type=subscript,
    name={\ensuremath{c}},
    description={Collagen fibers}
}
\newglossaryentry{wallLumenRatio}{
    type=latin,
    name={\ensuremath{\mathit{WLR}}},
    description={Wall to lumen ratio},
    unit={\si{-}}
}

\newglossaryentry{fsiInterface}{
    type=greek,
    name={\ensuremath{\gls{domainBoundary}^{\gls{fs}}}},
    description={Fluid-solid interface  = $\Gamma^f \cap \Gamma^s$},
    unit={-}
}
\newglossaryentry{fsiInterfaceFluid}{
    type=greek,
    name={\ensuremath{\gls{domainBoundary}_{fluid}^{\gls{fs}}}},
    description={%
        Fluid mesh boundary corresponding to the fluid-solid interface
    },
    unit={-}
}
\newglossaryentry{fsiInterfaceSolid}{
    type=greek,
    name={\ensuremath{\gls{domainBoundary}_{solid}^{\gls{fs}}}},
    description={%
        Solid mesh boundary corresponding to the fluid-solid interface
    },
    unit={-}
}
\newglossaryentry{fsiTolerance}{
    type=greek,
    name={\ensuremath{{\epsilon}_0}},
    description={Tolerance of the FSI coupling iterations},
    unit={-}
}
\newglossaryentry{fluidSolver}{
    type=latin,
    name={\ensuremath{\bm{\mathcal{F}}}},
    description={Symbolic representation of the fluid solver},
    unit={-}
}
\newglossaryentry{solidSolver}{
    type=latin,
    name={\ensuremath{\bm{\mathcal{S}}}},
    description={Symbolic representation of the solid solver},
    unit={-}
}
\newglossaryentry{addedMassOperator}{
    type=latin,
    name={\ensuremath{\bm{\mathcal{M}_{\mathrm{a}}}}},
    description={Symbolic representation of the added-mass operator},
    unit={-}
}
\newglossaryentry{addedMassEigenvalue}{
    type=latin,
    name={\ensuremath{\mu_\mathrm{a}}},
    description={Eigenvalue of the added-mass operator},
    unit={-}
}
\newglossaryentry{fsiTraction}{
    type=latin,
    name={\ensuremath{\cvec{y}}},
    description={%
        Solid solver input: traction exerted by the flow on the FSI interface
    },
    unit={\si{\pascal}}
}
\newglossaryentry{fsiInterfaceDispl}{
    type=latin,
    name={\ensuremath{\cvec{d}}},
    description={Position of the FSI interface},
    unit={\si{\meter}}
}
\newglossaryentry{fsDisplOutput}{
    type=latin,
    name={\ensuremath{\cvec{\tilde{d}}}},
    description={FSI interface position after a coupling iteration},
    unit={\si{\meter}}
}
\newglossaryentry{fsiResidualOperator}{
    type=latin,
    name={\ensuremath{\bm{\mathcal{R}}}},
    description={%
        Residual operator resulted from the fixed-point formulation of the FSI
        problem
    },
    unit={\si{\meter}}
}
\newglossaryentry{fsiInterfaceDoF}{
    type=latin,
    name={\ensuremath{N^{\gls{fs}}}},
    description={Number of degrees of freedom on the FSI interface},
    unit={-}
}
\newglossaryentry{fsiResidual}{
    type=latin,
    name={\ensuremath{\cvec{r}}},
    description={Discrete residual vector of the coupling iterations},
    unit={\si{\meter}}
}
\newglossaryentry{fsiReuse}{
    type=latin,
    name={\ensuremath{q}},
    description={Number of time-steps reused in the IQN-ILS algorithm},
    unit={-}
}

\newglossaryentry{aneurysm}{
    type=subscript,
    name={\ensuremath{ia}},
    description={%
        Indicates variables defined or integrated on the aneurysm surface
    }
}
\newglossaryentry{convexHull}{
    type=subscript,
    name={\ensuremath{ch}},
    description={%
        Indicates variables defined or integrated on the aneurysm surface
        convex hull
    }
}
\newglossaryentry{ostium}{
    type=subscript,
    name={\ensuremath{o}},
    description={%
        Indicates variables defined or integrated on the aneurysm ostium
        surface
    }
}

\newglossaryentry{domeHeight}{
    type=latin,
    name={\ensuremath{h_d}},
    description={%
        Maximum height of an aneurysm's dome perpendicular to ostium
        surface
    },
    unit={\si{\meter}}
}
\newglossaryentry{maximumDomeHeight}{
    type=latin,
    name={\ensuremath{h_{max}}},
    description={%
        Maximum height of an aneurysm's dome, measured from the ostium center
    },
    unit={\si{\meter}}
}
\newglossaryentry{bulgeHeight}{
    type=latin,
    name={\ensuremath{h_b}},
    description={%
        Bulge height of an aneurysm dome, i.e. the distance from the
        neck to the section with largest diameter
    },
    unit={\si{\meter}}
}
\newglossaryentry{neckDiameter}{
    type=latin,
    name={\ensuremath{d_n}},
    description={Diameter of an aneurysm's neck},
    unit={\si{\meter}}
}
\newglossaryentry{domeDiameter}{
    type=latin,
    name={\ensuremath{d_d}},
    description={Diameter of an aneurysm dome},
    unit={\si{\meter}}
}
\newglossaryentry{aspectRatio}{
    type=latin,
    name={\ensuremath{\mathit{AR}}},
    description={%
        Aspect ratio of an aneurysm, defined as the ratio between dome height
        and neck diameter
    },
    unit={-}
}
\newglossaryentry{meanBloodFlowRate}{
    type=latin,
    name={\ensuremath{\bar{q}_a}},
    description={Mean blood flow rate through an artery section},
    unit={\si{\milli\litre\per\second}}
}
\newglossaryentry{timeBloodFlowRate}{
    name={\ensuremath{q_p}},
    description={%
        Blood flow rate profile through an artery section of a specific patient
    },
    unit={\si{\milli\litre\per\second}}
}
\newglossaryentry{normBloodFlowRate}{
    name={\ensuremath{q^{*}}},
    description={%
        Normalized blood flow rate profile through the ICA section measured
        experimentally for a population.
    },
    unit={-}
}
\newglossaryentry{arteryDiameter}{
    type=latin,
    name={\ensuremath{d_a}},
    description={Diameter of an aneurysm's parent artery section},
    unit={\si{\meter}}
}

\newglossaryentry{bottleneckFactor}{
    type=latin,
    name={\ensuremath{\mathit{BF}}},
    description={Bottleneck factor},
    unit={-}
}
\newglossaryentry{bulgeLocation}{
    type=latin,
    name={\ensuremath{\mathit{BL}}},
    description={Bulge location},
    unit={-}
}
\newglossaryentry{sizeRatio}{
    type=latin,
    name={\ensuremath{\mathit{SR}}},
    description={Size ratio},
    unit={-}
}
\newglossaryentry{convexityRatio}{
    type=latin,
    name={\ensuremath{\mathit{CR}}},
    description={Convexity ratio},
    unit={-}
}
\newglossaryentry{isoperimetricRatio}{
    type=latin,
    name={\ensuremath{\mathit{IPR}}},
    description={Isoperimetric ratio},
    unit={-}
}
\newglossaryentry{undulationIndex}{
    type=latin,
    name={\ensuremath{\mathit{UI}}},
    description={Undulation index},
    unit={-}
}
\newglossaryentry{ellipticityIndex}{
    type=latin,
    name={\ensuremath{\mathit{EI}}},
    description={Ellipticity index},
    unit={-}
}
\newglossaryentry{nonsphericityIndex}{
    type=latin,
    name={\ensuremath{\mathit{NSI}}},
    description={Nonsphericity index},
    unit={-}
}
\newglossaryentry{conicityParameter}{
    type=latin,
    name={\ensuremath{\mathit{CP}}},
    description={Conicity Parameter},
    unit={-}
}
\newglossaryentry{areaAvgGaussianCurvature}{
    type=latin,
    name={\ensuremath{\mathit{GAA}}},
    description={Area-averaged Gaussian curvature of a surface},
    unit={\si{\per\square\milli\meter}}
}
\newglossaryentry{areaAvgMeanCurvature}{
    type=latin,
    name={\ensuremath{\mathit{MAA}}},
    description={Area-averaged mean curvature of a surface},
    unit={\si{\per\milli\meter}}
}
\newglossaryentry{l2NormGaussianCurvature}{
    type=latin,
    name={\ensuremath{\mathit{GLN}}},
    description={L2-norm of the Gaussian curvature field of a surface},
    unit={-}
}
\newglossaryentry{l2NormMeanCurvature}{
    type=latin,
    name={\ensuremath{\mathit{MLN}}},
    description={L2-norm of the mean curvature of a surface},
    unit={-}
}

\newglossaryentry{wallShearStressVector}{
    type=greek,
    name={\ensuremath{ \vec{\tau}_w }},
    description={%
        Wall shear stress vector defined on a vessel and aneurysm surface
    },
    unit={\si{\pascal}}
}
\newglossaryentry{wallShearStressMag}{
    type=greek,
    name={\ensuremath{{\tau}_w}},
    description={%
        Wall shear stress magnitude defined on a vessel and aneurysm surface
    },
    unit={\si{\pascal}}
}
\newglossaryentry{timeAvgWallShearStress}{
    type=latin,
    name={\ensuremath{\mathit{TAWSS}}},
    description={%
        Time-average WSS field
    },
    unit={\si{\pascal\per\meter}}
}

\newglossaryentry{wallShearStressGradient}{
    type=latin,
    name={\ensuremath{\vec{G}}},
    description={%
        Wall shear stress gradient vector, defined on the surface
        $pq$-coordinate system
    },
    unit={\si{\pascal\per\meter}}
}
\newglossaryentry{timeAvgWallShearStressGradient}{
    type=latin,
    name={\ensuremath{\mathit{TAWSSG}}},
    description={%
        Surface gradient of the time-averaged wall shear stress gradient
        vector, defined on the surface $p$-coordinate direction (the flow
        direction)
    },
    unit={\si{\pascal\per\meter}}
}
\newglossaryentry{avgShearDirectionVector}{
    type=latin,
    name={\ensuremath{\vec{p}}},
    description={$p$-coordinate direction (the flow direction)},
    unit={-}
}
\newglossaryentry{transShearDirectionVector}{
    type=latin,
    name={\ensuremath{\vec{q}}},
    description={%
        $q$-coordinate direction -- transversal to average flow direction
    },
    unit={-}
}
\newglossaryentry{oscillatoryShearIndex}{
    type=latin,
    name={\ensuremath{\mathit{OSI}}},
    description={The oscillatory shear index},
    unit={-}
}
\newglossaryentry{relativeResidenceTime}{
    type=latin,
    name={\ensuremath{\mathit{RRT}}},
    description={The relative residence time indicator},
    unit={-}
}
\newglossaryentry{gradientOscillatoryNumber}{
    type=latin,
    name={\ensuremath{\mathit{GON}}},
    description={The gradient oscillatory number},
    unit={-}
}
\newglossaryentry{aneurysmFormationIndicator}{
    type=latin,
    name={\ensuremath{\mathit{AFI}}},
    description={The \textit{potential} aneurysm formation indicator},
    unit={-}
}
\newglossaryentry{cardiacPeriod}{
    type=latin,
    name={\ensuremath{T_c}},
    description={Cardiac period},
    unit={\si{\second}}
}
\newglossaryentry{lowShearArea}{
    type=latin,
    name={\ensuremath{\mathit{lsa}}},
    description={Normalized area of low WSS defined on the aneurysm surface},
    unit={-}
}
\newglossaryentry{avgParentArteryWSS}{
    type=latin,
    name={\ensuremath{\avg{\tau}_{w,pa}}},
    description={%
        Averaged parent artery WSS, defined in the anterior portion of the
        parent artery
    },
    unit={\si{\pascal}}
}
\newglossaryentry{lowWSSReference}{
    type=latin,
    name={\ensuremath{(\gls{wallShearStressMag})_{low}}},
    description={Low WSS reference},
    unit={-}
}
\newglossaryentry{wssPulsatilityIndex}{
    type=latin,
    name={\ensuremath{\mathit{WSSPI}}},
    description={Pulsatility index of the wall shear stress magnitude},
    unit={-}
}
\newglossaryentry{wssTemporalGradient}{
    type=latin,
    name={\ensuremath{\mathit{WSSTG}}},
    description={%
        Maximal temporal derivative of the wall shear stress magnitude
    },
    unit={\si{\pascal\per\second}}
}
\newglossaryentry{transWallShearStress}{
    type=latin,
    name={\ensuremath{\mathit{transWSS}}},
    description={%
        Temporal average of the wall shear stress vector along its average
        transversal direction
    },
    unit={\si{\pascal}}
}
\newglossaryentry{speedup}{
    type=latin,
    name={\ensuremath{s}},
    description={Speedup},
    unit={-}
}
\newglossaryentry{residualVector}{
    type=latin,
    name={\ensuremath{\cvec{r}}},
    description={%
        Residual vector of a linear system $\matrix{A} \cvec{x} = \cvec{b}$
    },
    unit={-}
}
\newglossaryentry{residualNorm}{
    type=hidden,
    name={\ensuremath{R_s}},
    description={%
        Scaled normalized residual of a linear system in an iterative solver
    },
    unit={-}
}
\newglossaryentry{scaleFactor}{
    type=latin,
    name={\ensuremath{F_s}},
    description={Scale factor used for scaling the residual norm},
    unit={-}
}
\newglossaryentry{identityMatrix}{
    type=hidden,
    name={\ensuremath{\matrix{I}}},
    description={Identity matrix of $\realnumbers^{N \times N}$},
    unit={-}
}

\newacronym[category=standard,type=hidden]{1d}{1D}{one-dimensional}
\newacronym[category=standard,type=hidden]{2d}{2D}{two-dimensional}
\newacronym[category=standard,type=hidden]{3d}{3D}{three-dimensional}
\newacronym[category=standard,type=hidden]{4d}{3D}{four-dimensional}

\newacronym{cae}{CAE}{Computer-Aided Engineering}
\newacronym{fea}{FEA}{Finite Element Analysis}
\newacronym{cem}{CEM}{Computational Electromagnetic}
\newacronym{mbd}{MBD}{Multi-body Dynamics}

\newacronym{pde}{PDE}{partial differential equations}
\newacronym[category=never-short,type=hidden]{cfd}{CFD}{Computational Fluid Dynamics}
\newacronym{csd}{CSD}{Computational Solid Dynamics}
\newacronym{fdm}{FDM}{Finite Difference Method}
\newacronym{fem}{FEM}{Finite Element Method}
\newacronym[category=never-short,type=hidden]{fvm}{FVM}{Finite Volume Method}
\newacronym[type=hidden]{ale}{ALE}{arbitrary Lagrangian-Eulerian}
\newacronym{iqn-ils}{IQN-ILS}{Interface Quasi-Newton-Implicit Jacobian Least-Squares}
\newacronym{simple}{SIMPLE}{Semi-Implicit Method for Pressure Linked Equations}
\newacronym[category=only-short,type=hidden]{piso}{PISO}{Pressure Implicit with Splitting Operators}
\newacronym{fsi}{FSI}{Fluid-Solid Interaction}

\newacronym[category=never-short,type=hidden]{bc}{BC}{boundary condition}
\newacronym{am}{AM}{added-mass}
\newacronym{dnf}{DNF}{Dirichlet-Neumann formulation}
\newacronym{dof}{DOF}{degrees of freedom}

\newacronym{ia}{IA}{intracranial aneurysm}
\newacronym{aaa}{AAA}{abdominal aortic aneurysm}
\newacronym[category=standard,type=hidden]{vmtk}{VMTK}{Vascular Modeling Toolkit}
\newacronym[category=never-short,type=hidden]{dsa}{DSA}{digital subtraction angiography}
\newacronym{cta}{CTA}{computational tomography angiography}
\newacronym{wlr}{WLR}{wall-to-lumen ratio}
\newacronym[category=never-short,type=hidden]{pae}{PAE}{peri-aneurysmal environment}
\newacronym{sah}{SAH}{subarachnoid hemorrhage}
\newacronym[category=standard,type=hidden]{vtk}{VTK}{Visualization Toolkit}

\newacronym{wss}{WSS}{wall shear stress}
\newacronym{pswss}{PSWSS}{peak systole wall shear stress}
\newacronym{tawss}{TAWSS}{time-averaged wall shear stress}
\newacronym{tawssg}{TAWSSG}{time-averaged WSS gradient}
\newacronym{wsstg}{WSSTG}{maximum wall shear stress temporal gradient}
\newacronym{wsssg}{WSSSG}{wall shear stress surface gradient}
\newacronym{wsspi}{WSSPI}{wall shear stress pulsatility index}
\newacronym{transwss}{transWSS}{transversal wall shear stress}
\newacronym{osi}{OSI}{oscillatory shear index}
\newacronym{rrt}{RRT}{relative residence time}
\newacronym{afi}{AFI}{aneurysm formation indicator}
\newacronym{gon}{GON}{gradient oscillatory number}
\newacronym{lsa}{LSA}{low shear area}

\newacronym{mca}{MCA}{middle cerebral artery}
\newacronym{ica}{ICA}{internal carotid artery}
\newacronym{aca}{ACA}{anterior cerebral artery}
\newacronym{acoa}{ACoA}{anterior communicating artery}
\newacronym{pica}{PICA}{posterior inferior cerebellar artery}
\newacronym{pca}{PCA}{posterior cerebral artery}
\newacronym{ba}{BA}{basilar artery}
\newacronym{va}{VA}{vertebral arteries}
\newacronym{oa}{OA}{ophthalmic artery}

\newacronym[type=hidden]{fapesp}{FAPESP}{São Paulo Research Foundation}
\newacronym[type=hidden]{unesp}{UNESP}{São Paulo State University}

\newacronym{1wfsi}{1WFSI}{one-way fluid-solid interaction}
\newacronym[category=never-short,type=hidden]{2wfsi}{2WFSI}{two-way fluid-solid interaction}
\newacronym[category=never-short,type=hidden]{hwp}{HWP}{hemodynamic wall parameter}

\newacronym[category=never-short,type=hidden]{mr}{MR}{Mooney-Rivlin}
\newacronym{ul}{UL}{updated Lagrangian}
\newacronym{tl}{TL}{total Lagrangian}
\newacronym[category=never-short,type=hidden]{svk}{SVK}{St.\ Venant-Kirchhoff}

\newacronym[category=never-short,type=hidden]{mri}{MRI}{Magnetic Resonance Imaging}
\newacronym[category=never-short,type=hidden]{uwm}{UWM}{uniform-wall model}
\newacronym[category=never-short,type=hidden]{awm}{AWM}{abnormal-wall model}
\newacronym[category=never-short,type=hidden]{sd}{SD}{standard deviation}

\newacronym[type=hidden]{case50}{rMCA1}{}
\newacronym[type=hidden]{case23}{rMCA2}{}
\newacronym[type=hidden]{case37}{rICA1}{}
\newacronym[type=hidden]{case1}{rMCA3}{}
\newacronym[type=hidden]{case4ruptured}{rICA2}{}
\newacronym[type=hidden]{case2}{rICA3}{}

\newacronym[type=hidden]{case51}{urMCA1}{}
\newacronym[type=hidden]{case74}{urICA1}{}
\newacronym[type=hidden]{case4unruptured}{urMCA2}{}
\newacronym[type=hidden]{case13}{urICA2}{}
\newacronym[type=hidden]{case8}{urMCA3}{}
\newacronym[type=hidden]{case64}{urMCA4}{}
\newacronym[type=hidden]{case11}{urMCA5}{}

\newacronym[type=hidden]{m1}{\ensuremath{M_1}}{}
\newacronym[type=hidden]{m2}{\ensuremath{M_2}}{}
\newacronym[type=hidden]{m3}{\ensuremath{M_3}}{}

\newacronym[type=hidden]{dt1}{\ensuremath{\gls{deltat}_1}}{}
\newacronym[type=hidden]{dt2}{\ensuremath{\gls{deltat}_2}}{}
\newacronym[type=hidden]{dt3}{\ensuremath{\gls{deltat}_3}}{}

\newcommand{\case}[1]{\acrshort{#1}}

\makeatletter
\newcommand*{\Xbar}{}%
\DeclareRobustCommand*{\Xbar}{%
  \mathpalette\@Xbar{}%
}
\newcommand*{\@Xbar}[2]{%
  \sbox0{$#1\mathrm{X}\m@th$}%
  \sbox2{$#1X\m@th$}%
  \rlap{%
    \hbox to\wd2{%
      \hfill
      $\overline{%
        \vrule width 0pt height\ht0 %
        \kern\wd0 %
      }$%
    }%
  }%
  \copy2 %
}
\makeatother

\usepackage{xifthen}% provides \isempty test
\newcommand{\isoinvariant}[2]{%
    \ifthenelse{\isempty{#2}}%
        {\ensuremath{I^{*}_{#1}}}% if #2 is empty
        {\ensuremath{I^{*}_{#1,\gls{#2}}}}% if #2 is not empty
}

\newcommand{\materialcoeff}[3]{%
    \ifthenelse{\isempty{#3}}%
        {\ensuremath{\gls{#1}_{#2}}}%       if #3 is empty
        {\ensuremath{\gls{#1}_{#2#3}}}% if #3 is not empty
}

\newglossaryentry{VonMisesStress}{
    type=latin,
    name={\ensuremath{\gls{CauchyStressComp}_{0}}},
    description={Von Mises or equivalent stress},
    unit={\si{\pascal}}
}
\newglossaryentry{maxPrincipalCauchyStress}{
    type=latin,
    name={\ensuremath{\gls{CauchyStressComp}_{1}}},
    description={Largest principal eigenvalue of the Cauchy stress tensor},
    unit={\si{\pascal}}
}
\newglossaryentry{maxStretch}{
    type=latin,
    name={\ensuremath{\gls{stretch}_{1}}},
    description={Largest eigenvalue of the stretch tensor},
    unit={\si{\pascal}}
}

\newglossaryentry{penaltyParameter}{
    type=greek,
    name={\ensuremath{\kappa}},
    description={Penalty parameter},
    unit={\si{\pascal}}
}

\newglossaryentry{pressureDiffusivity}{
    type=greek,
    name={\ensuremath{\gamma}},
    description={%
        Diffusivity of the Rhie-Chow correction in the pressure
        equation
    },
    unit={\si{\pascal}}
}

\newglossaryentry{modifiedCompressibility}{
    type=greek,
    name={\ensuremath{\gls{isothermalCompressibility}^{\gls{fluid}^{*}}}},
    description={%
        Modified isothermal compressibility of a weakly compressible fluid
    },
    unit={\si{\square\second\per\square\meter}}
}

\newglossaryentry{MurnaghanCoeff}{
    type=latin,
    name={\ensuremath{n^{\gls{fluid}}}},
    description={Material constant defined in the Murnaghan equation of state},
    unit={-}
}

\newglossaryentry{referenceDensity}{
    type=greek,
    name={\ensuremath{ \gls{density}^{\gls{fluid}}_{\mathrm{ref}}}},
    description={Reference density of the weakly compressible model},
    unit={\si{\kilo\gram\per\cubic\meter}}
}

\newglossaryentry{vasculature}{
    type=subscript,
    name={\ensuremath{v}},
    description={%
        Indicates variables defined or integrated on the vasculature surface
    }
}
\newglossaryentry{parentArtery}{
    type=subscript,
    name={\ensuremath{pa}},
    description={%
        Indicates variables defined or integrated on the parent artery surface
        of an aneurysm
    }
}
\newglossaryentry{branch}{
    type=subscript,
    name={\ensuremath{b}},
    description={%
        Variables or properties of the vascular branches
    }
}

\newglossaryentry{peakSystole}{
    type=subscript,
    name={\ensuremath{ps}},
    description={%
        Indicates variables defined at the peak systole instant
    }
}
\newglossaryentry{lowDiastole}{
    type=subscript,
    name={\ensuremath{ld}},
    description={%
        Indicates variables defined at the low diastole instant
    }
}

\newglossaryentry{uniformWallThickness}{
    type=latin,
    name={\ensuremath{{e_{uw}}}},
    description={Uniform thickness of the artery walls},
    unit={\si{\milli\meter}}
}

\newglossaryentry{wallThickness}{
    type=latin,
    name={\ensuremath{{e_w}}},
    description={%
        Thickness field of a vascular wall (i.e. aneurysms and branches)
    },
    unit={\si{\milli\meter}}
}
\newglossaryentry{branchesThickness}{
    type=latin,
    name={\ensuremath{{e_{\gls{branch}}}}},
    description={Thickness field of the branches walls},
    unit={\si{\milli\meter}}
}
\newglossaryentry{aneurysmThickness}{
    type=latin,
    name={\ensuremath{e_{\gls{aneurysm}}}},
    description={Thickness field of an aneurysm's wall},
    unit={\si{\milli\meter}}
}

\newglossaryentry{aneurysmThicknessFactor}{
    type=latin,
    name={\ensuremath{f_a}},
    description={%
        Scale factor used on the calculation of the aneurysm thickness
    },
    unit={\si{-}}
}
\newglossaryentry{abnormalHemodynamics}{
    type=subscript,
    name={\ensuremath{ab}},
    description={%
        Properties of an aneurysm wall due to abnormal hemodynamics
    },
}

\newglossaryentry{abnormalThicknessFactor}{
    type=latin,
    name={\ensuremath{f_{\gls{abnormalHemodynamics}}}},
    description={%
        Scale factor used on the calculation of the aneurysm thickness on
        abnormal hemodynamics patches
    },
    unit={\si{-}}
}

\newglossaryentry{aneurysmPulsatility}{
    type=greek,
    name={\ensuremath{\delta_{v}}},
    description={Pulsatility of an intracranial aneurysm},
    unit={-}
}

\newglossaryentry{distanceToNeckLine}{
    type=latin,
    name={\ensuremath{g_n}},
    description={%
        Geodesic distance of the vasculature to the aneurysm neck line
    },
    unit={\si{\milli\meter}}
}

\newglossaryentry{lumenDiameter}{
    type=latin,
    name={\ensuremath{d_l}},
    description={%
        Vasculature lumen diameter at a specified section
    },
    unit={\si{\milli\meter}}
}

\newglossaryentry{intracranialPressure}{
    type=latin,
    name={\ensuremath{\gls{pressure}_{o}}},
    description={Intracranial pressure},
    unit={\si{\pascal}}
}

\newglossaryentry{SpearmanCoeff}{
    type=greek,
    name={\ensuremath{\rho_{\mathrm{S}}}},
    description={Spearman's correlation coefficient},
    unit={-}
}
\newglossaryentry{PearsonCoeff}{
    type=greek,
    name={\ensuremath{\rho_{\mathrm{P}}}},
    description={Pearson's correlation coefficient},
    unit={-}
}

\newglossaryentry{HookeanCauchyStress}{
    type=greek,
    name={\ensuremath{{\gls{CauchyStress}}_{H}}},
    description={Cauchy stress tensor of the Hookean materal model},
    unit={\si{\pascal}}
}

\newglossaryentry{HookeanCauchyStressComp}{
    type=hidden,
    name={\ensuremath{{\gls{CauchyStressComp}}_{H}}},
    description={Cauchy stress tensor component of the Hookean material model},
    unit={\si{\pascal}}
}

\newglossaryentry{cylinderInnerRadius}{
    type=latin,
    name={\ensuremath{R_i}},
    description={Cylinder inner radius},
    unit={\si{\meter}}
}

\newglossaryentry{cylinderOuterRadius}{
    type=latin,
    name={\ensuremath{R_o}},
    description={Cylinder outer radius},
    unit={\si{\meter}}
}
\newglossaryentry{transmuralPressure}{
    type=latin,
    name={\ensuremath{\gls{pressure}_{tr}}},
    description={Transmural pressure},
    unit={\si{\pascal}}
}
\newglossaryentry{firstPrincipalCurvature}{
    type=greek,
    name={\ensuremath{\kappa_{1}}},
    description={First principal curvature of a surface},
    unit={\si{\per\meter}}
}
\newglossaryentry{secondPrincipalCurvature}{
    type=greek,
    name={\ensuremath{\kappa_{2}}},
    description={Second principal curvature of a surface},
    unit={\si{\per\meter}}
}
\newglossaryentry{meanCurvature}{
    type=latin,
    name={\ensuremath{H}},
    description={Mean curvature of a surface},
    unit={\si{\per\meter}}
}
\newglossaryentry{GaussianCurvature}{
    type=latin,
    name={\ensuremath{K}},
    description={Gaussian curvature of a surface},
    unit={\si{\per\square\meter}}
}

\newglossaryentry{membraneTension}{
    type=latin,
    name={\ensuremath{T}},
    description={Membrane tension},
    unit={\si{\pascal\meter}}
}
\newglossaryentry{volumetric}{
    type=subscript,
    name={\ensuremath{vol}},
    description={%
        Indicates the volumetric part of the strain-energy function
    }
}
\newglossaryentry{isochoric}{
    type=subscript,
    name={\ensuremath{iso}},
    description={%
        Indicates the isochoric or volume-preserving part of the strain-energy
        function
    }
}
\newglossaryentry{materialProjectionTensor}{
    type=latin,
    name={\ensuremath{\fourthTensor{P}}},
    description={%
        Projection tensor that maps the isochoric stress tensors to the
        physical stress tensor in Lagrangian coordinates
    },
    unit={-}
}
\newglossaryentry{YeohCoeffs}{
    type=hidden,
    name={\ensuremath{c}},
    description={%
        Material coefficients symbol of Yeoh's strain energy function
    },
    unit={\si{\pascal}}
}
\newglossaryentry{FungCoeffs}{
    type=hidden,
    name={\ensuremath{k}},
    description={%
        Material coefficients symbol of Fung's isotropic strain energy function
        ($c_1$ is given in \si{\pascal} whereas the others are dimensioneless)
    },
    unit={\si{\pascal}}
}
\newglossaryentry{axialWaveSpeed}{
    type=hidden,
    name={\ensuremath{U_n}},
    description={%
        Axial or longitudinal wave speed used in the advective boundary
        condition
    },
    unit={\si{\meter\per\second}}
}
\newglossaryentry{genericRelativeDifference}{
    type=hidden,
    name={\ensuremath{D}},
    description={%
        Relative difference between a wall mechanics metric between different
        modeling choices
    },
    unit={-}
}
\newglossaryentry{genericAbsoluteDifference}{
    type=hidden,
    name={\ensuremath{D}},
    description={%
        Absolute difference between a wall mechanics metric between different
        modeling choices
    },
    unit={-}
}
\newglossaryentry{meshCellsNumber}{
    type=hidden,
    name={\ensuremath{N_M}},
    description={%
        Number of cells in a computational mesh
    },
    unit={-}
}
\newglossaryentry{meshSurfaceCellDensity}{
    type=hidden,
    name={\ensuremath{\rho_{S}}},
    description={%
        Cells surface density in a computational mesh
    },
    unit={\si{cells\per\square\milli\meter}}
}
\newglossaryentry{maxEdgeSize}{
    type=hidden,
    name={\ensuremath{l_e}},
    description={%
        Maximum edge length size in a computational mesh
    },
    unit={\si{\milli\meter}}
}
\newglossaryentry{meshRefinementRatio}{
    type=latin,
    name={\ensuremath{r}},
    description={%
        Ration between element size of systematically refined meshes
    },
    unit={-}
}
\newglossaryentry{fvmOrder}{
    type=latin,
    name={\ensuremath{p}},
    description={%
        Formal order of accuracy of the Finite Volume Method
    },
    unit={-}
}
\newglossaryentry{discretizationError}{
    type=latin,
    name={\ensuremath{e}},
    description={%
        Discretization error of a particular mesh computed using Richardson's
        extrapolation
    },
    unit={-}
}
\newglossaryentry{pValue}{
    type=hidden,
    name={\ensuremath{p}},
    description={P-value of statistical tests},
    unit={-}
}
\newglossaryentry{significanceLevel}{
    type=hidden,
    name={\ensuremath{\alpha}},
    description={Significance level for statistical tests},
    unit={-}
}
\newglossaryentry{standardDeviation}{
    type=hidden,
    name={\ensuremath{\mathit{SD}}},
    description={Standard-deviation of a sample's distribution},
    unit={-}
}
\newglossaryentry{ruptureProbability}{
    type=latin,
    name={\ensuremath{P_{r}^{\textsc{ia}}}},
    description={Rupture probability of an intracranial aneurysm},
    unit={-}
}
\newglossaryentry{Yeoh}{
    type=superscript,
    name={\ensuremath{\mathrm{Y}}},
    description={%
        Indicates variables defined for the Yeoh hyperelastic law
    }
}
\newglossaryentry{Fung}{
    type=superscript,
    name={\ensuremath{\mathrm{F}}},
    description={%
        Indicates variables defined for the Fung hyperelastic law
    }
}
\newglossaryentry{MooneyRivlin}{
    type=superscript,
    name={\ensuremath{\mathrm{MR}}},
    description={%
        Indicates variables defined for the Mooney-Rivlin hyperelastic law
    }
}
\newglossaryentry{lumen}{
    type=superscript,
    name={\ensuremath{\mathrm{l}}},
    description={%
        Indicates variables defined on the lumen surface
    }
}
\newglossaryentry{ablumen}{
    type=superscript,
    name={\ensuremath{\mathrm{abl}}},
    description={%
        Indicates variables defined on the ablumen surface
    }
}
\newglossaryentry{mid}{
    type=superscript,
    name={\ensuremath{\mathrm{m}}},
    description={%
        Indicates variables defined on the mid surface of the vessel
    }
}
\newglossaryentry{iaWallMetric}{
    type=latin,
    name={\ensuremath{\mathcal{M}}},
    description={%
        Generic representation of a metric of the stress or stretch fields over
        the wall of an aneurysm
    }
}
\newglossaryentry{rupturedGroup}{
    type=latin,
    name={\ensuremath{\mathcal{R}}},
    description={%
        The group of ruptured aneurysms
    }
}
\newglossaryentry{unrupturedGroup}{
    type=latin,
    name={\ensuremath{\mathcal{U}}},
    description={%
        The group of unruptured aneurysms
    }
}
\newglossaryentry{typeIPatch}{
    type=subscript,
    name={\ensuremath{I}},
    description={%
        Indicates variables defined on the type-I patches of the aneurysm sac
    }
}
\newglossaryentry{typeIIPatch}{
    type=subscript,
    name={\ensuremath{II}},
    description={%
        Indicates variables defined on the type-II patches of the aneurysm sac
    }
}
\newglossaryentry{normalPatch}{
    type=subscript,
    name={\ensuremath{R}},
    description={%
        Indicates variables defined on the normal patches of the aneurysm sac
        (i.e., not type I or type II)
    }
}
\newglossaryentry{neckPatch}{
    type=subscript,
    name={\ensuremath{N}},
    description={%
        Indicates the neck patch of an \gls{ia} sac, defined by the
        physician-oriented patching
    }
}
\newglossaryentry{bodyPatch}{
    type=subscript,
    name={\ensuremath{B}},
    description={%
        Indicates the body patch of an \gls{ia} sac, defined by the
        physician-oriented patching
    }
}
\newglossaryentry{domePatch}{
    type=subscript,
    name={\ensuremath{D}},
    description={%
        Indicates the dome patch of an \gls{ia} sac, defined by the
        physician-oriented patching
    }
}
\newglossaryentry{ellipticPatch}{
    type=subscript,
    name={\ensuremath{E}},
    description={%
        Elliptic patches of a surface
    }
}
\newglossaryentry{hyperbolicPatch}{
    type=subscript,
    name={\ensuremath{H}},
    description={%
        Hyperbolic patches of a surface
    }
}

    \makeglossaries

    \usepackage{chngcntr}

    \usepackage{nth}

    \usepackage[capitalise]{cleveref}

\begin{document}

    \begin{frontmatter}

    \title{%
         On the major role played by the curvature of intracranial aneurysms
         walls in determining their mechanical response, local hemodynamics,
         and rupture likelihood
    }

    \author[hiae-research]{I. L. Oliveira\corref{cor}}
    \ead{iago.oliveira@unesp.br}

    \author[ucd]{P. Cardiff}
    \ead{philip.cardiff@ucd.ie}

    \author[hiae]{C.E. Baccin}
    \ead{carlos.baccin@einstein.br}

    \author[einstein]{R.T. Tatit}
    \ead{rtrindadetatit@gmail.com}

    \author[feis-short]{J.L. Gasche}
    \ead{jose.gasche@unesp.br}

    \cortext[cor]{Corresponding author}

    %\fntext[fn1]{This is the specimen author footnote.}
    %\fntext[fn2]{Another author footnote, but a little longer.}
    %\fntext[fn3]{Yet another footnote. Indeed you can use any number of author footnotes.}

    \address[hiae-research]{%
        Postdoctoral Researcher, Centro de Pesquisa em Imagem, Hospital
        Israelita Albert Einstein, % São Paulo, Brazil\\
        Av. Albert Einstein, 627 -- 4º andar, Bloco D, Morumbi, São Paulo --
        SP, CEP: 05652-901, Brazil
    }

    \address[ucd]{%
        University College Dublin (UCD), School of Mechanical and
        Materials Engineering, Dublin, Ireland
    }

    \address[hiae]{%
        Interventional Neuroradiologist, Hospital Israelita Albert Einstein, %
        São Paulo, Brazil
    }

    \address[einstein]{%
        Albert Einstein Israeli Faculty of Health Sciences, %
        São Paulo, Brazil
    }

    \address[feis-short]{%
        São Paulo State University (UNESP), School of Engineering, %
        Mechanical Engineering Department
    }

    %\address[feis-long]{%
    %    São Paulo State University (UNESP), School of Engineering,
    %    Mechanical Engineering Department, Thermal Sciences Building,
    %    Avenida Brasil, 56, Ilha Solteira - SP, Brazil
    %}

    \begin{abstract}
        The properties of \glspl{ia} walls are known to be driven by the
        underlying hemodynamics adjacent to the \gls{ia} sac. Different
        pathways exist explaining the connections between  hemodynamics and
        local tissue properties. The emergence of such theories is essential if
        one wishes to compute the mechanical response of a patient-specific
        \gls{ia} wall and predict its rupture. Apart from the hemodynamics and
        tissue properties, one could assume that the mechanical response also
        depends on the local morphology, more specifically, the wall curvature,
        with larger values at highly-curved wall portions. Nonetheless, this
        contradicts observations of \gls{ia} rupture sites more often found at
        the dome, where the curvature is lower. This seeming contradiction
        indicates a complex interaction between local hemodynamics, wall
        morphology, and mechanical response, which warrants further
        investigation. This was the main goal of this work.  We accomplished
        this by analysing the stress and stretch fields in different regions of
        the wall for a sample of \glspl{ia}, which have been classified based
        on particular local hemodynamics and local curvature. Pulsatile
        numerical simulations were performed using the one-way fluid-solid
        interaction strategy implemented in OpenFOAM (solids4foam toolbox). We
        found that the variable best correlated with regions of high stress and
        stretch was the wall curvature. Additionally, our data suggest a
        connection between the local curvature and local hemodynamics,
        indicating that the curvature is a property that could be used to
        assess both mechanical response and hemodynamic conditions, and,
        moreover, to suggest new metrics based on the curvature to predict
        the likelihood of rupture.
    \end{abstract}


\end{frontmatter}

    \sisetup{%
        detect-all,
        output-decimal-marker={.},
        multi-part-units=single,
        separate-uncertainty=true,
        per-mode=symbol,
    }%

    \glsresetall

        \section{Introduction} \label{sec:introduction}

    Based on radiographic and autopsy series, it is estimated that the
    prevalence of \glspl{ia} is \SI{3.2}{\percent} in a population without
    comorbidity \citep{Vlak2011}. Most of these \glspl{ia} will not rupture, as
    rupture rates are relatively low \citep{Brown2014}. However, the ones that
    do rupture may cause a poor outcome to the patient due to subarachnoid
    aneurysmal haemorrhages, which are the main complication caused by the
    rupture of \glspl{ia}. Subarachnoid aneurysmal haemorrhages occur with a
    frequency of \num{7.9} per \num{100000} person-years, according to
    \citet{Etminan2019}. Regarding their detection, apart from the cases of
    rupture and large lesions (because they tend to compress any surrounding
    structures), most aneurysms are detected incidentally when patients seek
    clinicians for other reasons. Fortunately, their detection has been
    increasing substantially \citep{Vlak2011} due to the higher frequency of
    cranial imaging and better imaging techniques \citep{Hacein-Bey2011}.
    Nonetheless, a clinician faces a difficult decision on whether to treat a
    recently-diagnosed \gls{ia} due to the delicate balance between the risk
    posed by the available treatments and the \gls{ia} rupture risk, which
    depends on its size and location \citep{ClaiborneJohnston2000, Brown2014}.
    Significant effort has been made in recent decades to find a better
    indicator of rupture to facilitate this decision; for example,
    machine-learning techniques have been proposed to decide whether a
    particular aneurysm will rupture, as a function of clinical, morphological
    and, in some cases, hemodynamics data of patient-specific \glspl{ia}
    \citep{RajabzadehOghaz2020a}.

    In early studies, aneurysms were thought to be possibly congenital
    \citep{Stehbens1989}, but today it is widely accepted that they are
    acquired lesions related to the interaction between the hemodynamic
    environment acting on the aneurysm luminal side and the wall tissue. In
    this context, the \gls{wss}, the tangential flow traction acting on the
    vessel lumen, is one of the most important parameters related to their
    development \citep{Fukazawa2015}. While the onset of an \gls{ia} is
    generally agreed to be caused by high \gls{wss} levels and high and
    positive spatial gradients of \gls{wss} \citep{Geers2017,Soldozy2019}, its
    subsequent growth is most likely multifactorial and currently explained by
    different pathways connecting particular hemodynamic environments and
    aneurysm enlargement \citep{Meng2014,Frosen2019}. These pathways have been
    correlated with particular wall phenotypes defined by morphological and
    mechanical features, such as wall thickening and stiffening, that may occur
    in the regions of an \gls{ia} wall where those adjacent hemodynamic
    environments occur \citep{Meng2014,Cebral2017}. Additionally, these
    mechanisms impact the strength of the aneurysm wall \citep{Robertson2015},
    possibly creating weaker regions. Then, from a mechanical point of view, a
    patient-specific \gls{ia} may rupture when the mechanical stress, also
    induced in the wall tissue by the \gls{wss} and pressure fields of the
    blood flow locally, exceeds the hemodynamics-driven wall strength.

    Nevertheless, the connections between the stresses and strains developed in
    an \gls{ia} wall, its hemodynamics environment and the rupture-site
    location are still poorly understood. For example, a non-intuitive fact
    about rupture-site locations is that only a small percentage are found at
    the neck \citep{Park2012}, where the stresses are expected to be higher due
    to the saddle nature of the wall curvature, while the majority of them
    occur on the aneurysm dome \citep{Kono2012,Zhang2016}.
    %Therefore, it is reasonable to assume that exists a complex exists among
    %the failure properties of the tissue, the stress levels induced by the
    %hemodynamic environment, and the morphology of the wall.

    \textit{In-vivo} and non-invasive methods to measure the mechanical
    response of arterial walls are still emerging \citep{Mahmoud2009}, mainly
    to measure the properties and motion of the aorta \citep{Danpinid2010}.
    Nonetheless, these methods may not be useful for accurate measurements of
    \gls{ia} wall motion, given their size. Moreover, techniques to measure the
    stresses \textit{in-vivo} are still challenging. Hence, most of the works
    on the subject have used numerical simulations
    \citep{Torii2010,Ramachandran2012,Lee2013a} to compute the stress and
    deformation of \gls{ia} walls, but with limiting modelling assumptions
    about the tissue \citep{Voss2016}. This is probably explained by the
    limited amount of information on the mechanical properties of
    patient-specific \glspl{ia}, which are difficult to be measure
    experimentally either with extracted tissue samples or \emph{in vivo}.
    This situation, though, is improving, with recent work demonstrating how to
    experimentally obtain the mechanical properties of patient-specific
    \gls{ia} tissue \citep{Costalat2011,Robertson2015}.

    The precise role played by the local morphology of the walls in the stress
    and strain levels is also poorly understood. \Gls{ia} walls may be assumed
    as thick shells, as intracranial arteries have an average thickness of
    about \SI{8.6}{\percent} of their local diameter \citep{Nakagawa2016}. For
    \glspl{ia}, since early studies, it was recognised that \enquote{from a
    mechanical perspective, shape and thickness are more important contributors
    to rupture-potential than overall size}, as noted by \citet{Humphrey2000}.
    From a theoretical perspective, this is understandable: Laplace's law for
    membranes predicts that the stress induced in a membrane is inversely
    proportional to its thickness and principal curvatures (shape is
    intrinsically related to local curvature). For thick shells, though,
    Laplace's laws do not hold locally \citep{Fung1993}, which was also
    confirmed for \gls{ia} walls by \citet{Ma2007} and
    \citet{Ramachandran2012}.

    Nevertheless, \gls{ia} sac curvature has been the subject of few works. We
    conducted a query in Web Of Science\R  for published papers with terms
    related to \enquote{\glsxtrlongpl{ia}} and \enquote{numerical simulation}
    and found only \SI{2.5}{\percent} of items also mentioned \enquote{surface
    curvature} or \enquote{surface shape} as their main topic. An even smaller
    percentage of works (\SI{0.32}{\percent}) was found when comparing papers
    that investigated \glspl{ia} in general, i.e. that did not necessarily
    involve numerical simulations. One of those few studies that employed
    numerical simulations, by \citet{Ma2007}, found a strong correlation
    between the maximum principal stress and high curvature regions, measured
    by the principal curvature. However, the authors recognise that the precise
    nature of this relationship should be further investigated.

    The goal of this article is to investigate the connections between the
    mechanical wall response, local hemodynamics, and morphological features of
    the wall of patient-specific \glspl{ia}, more specifically its curvature,
    by using ruptured and unruptured \glspl{ia} geometries. To predict the
    stress and strain fields in real aneurysms geometries, we employed a
    \glsxtrlong{1wfsi} numerical strategy implemented in the solids4foam
    toolbox \citep{Cardiff2018}, an extension of OpenFOAM\R \citep{Weller1998},
    with a realistic modeling of the walls and material constants. Different
    regions of the aneurysm walls were analysed and characterised by different
    hemodynamics and curvature properties.

        \section{Numerical Methodology} \label{sec:numericalMethodology}

    \subsection{Sample Selection and Geometry Preparation}
    \label{sec:sampleSelection}

    We selected twelve vascular geometries from \gls{dsa} examinations
    collected retrospectively and harbouring thirteen bifurcation \glspl{ia},
    all of them originating from the \gls{ica} and \gls{mca}. These images were
    segmented using the \gls{vmtk}\R library \citep{vmtk} with the level-sets
    method \citep{Piccinelli2009} by selecting a region of interest that
    encloses only the aneurysm and its surrounding vessels. Then, a
    triangulated surface was generated with the Marching Cubes algorithm
    \citep{Antiga2002, Antiga2008}, and inlet and outlet profiles were opened
    on it to impose \glspl{bc} for the numerical simulations. Seven \glspl{ia}
    were unruptured, and six were ruptured. In this text, we label these images
    by appending their rupture status, prefix \enquote{r} for ruptured and
    \enquote{ur} for unruptured, to their parent artery. For example, a
    ruptured case in the \gls{mca} bifurcation is labelled \enquote{rMCA},
    followed by a natural number in case of repetition. The use of those
    geometries was approved by the Research Ethics Committees of the Albert
    Einstein Israelite Hospital, São Paulo, Brazil, where nine of them were
    collected, and of the Faculty of Medicine of São Paulo State University
    (UNESP), Campus of Botucatu, Brazil. Due to the lack of sufficient ruptured
    cases in the original dataset, three additional vascular geometries were
    obtained from the Aneurisk dataset repository \citep{aneurisk} (available
    under the \enquote{CC BY-NC 3.0} license).

    \subsection{%
        Physical and Mathematical Modeling and Computational Strategies
    }
    \label{sec:mechanicalModeling}

    We employed the so-called \enquote{\gls{1wfsi}} strategy to numerically
    solve for the interaction between the blood flow and the solid wall motion
    \citep{Hirschhorn2020}. By using this technique, the pulsatile fluid flow
    is solved by assuming a rigid \gls{fsi} interface, i.e. with zero mass
    flux, zero pressure gradient, and the no-slip condition. Then, at each
    instant in time along the cardiac cycle, the traction on this interface due
    to the blood flow is transferred to the solid \gls{fsi} interface
    counterpart and applied as a traction \gls{bc}. The resulting solid
    deformation is not transferred back to the fluid domain. Blood was assumed
    to be a weakly compressible Newtonian fluid flowing in an isothermal
    laminar regime and the \gls{ia} and artery wall tissue was assumed as
    isotropic and represented by the hyperelastic three-parameters \gls{mr} law
    by using the pseudoelastic modeling approach \citep{Fung1979}. The details
    about the \gls{1wfsi} numerical strategy, the fluid and solid modelling,
    with their respective \glspl{bc}, can be found in  \citep{Oliveira2022}
    that employed the same modelling. The numerical simulations were performed
    in solids4foam \citep{Cardiff2018}, compiled with the foam-extend library
    \citep{foam-extend,Weller1998}, version 4.0 and the details about the
    discretisation techniques employed can also be found in
    \citet{Oliveira2022}, including mesh and time-step independence tests.

    The detailed modelling of the wall thickness, \gls{wallThickness}, and the
    material constants of the \gls{mr} law of the vascular walls was also
    presented in \citet{Oliveira2022} and, for reference, was labelled as the
    \enquote{uniform-wall model} in that work due to its context.
    Nevertheless, they deserve further comments here. These properties were
    modelled with a globally heterogeneous model over the whole vasculature by
    computing different thicknesses and material constants on the branches and
    the aneurysm sac. This modelling was chosen to eliminate the local
    influence of the thickness and material constant heterogeneity on the
    mechanical response of the aneurysm sac --- it is known that \gls{ia} walls
    are heterogenous \citep{Kadasi2013} --- while using a sufficiently
    realistic model of the walls.

\begin{figure}[!htb]
        \includegraphics[%
            width=\textwidth
        ]   {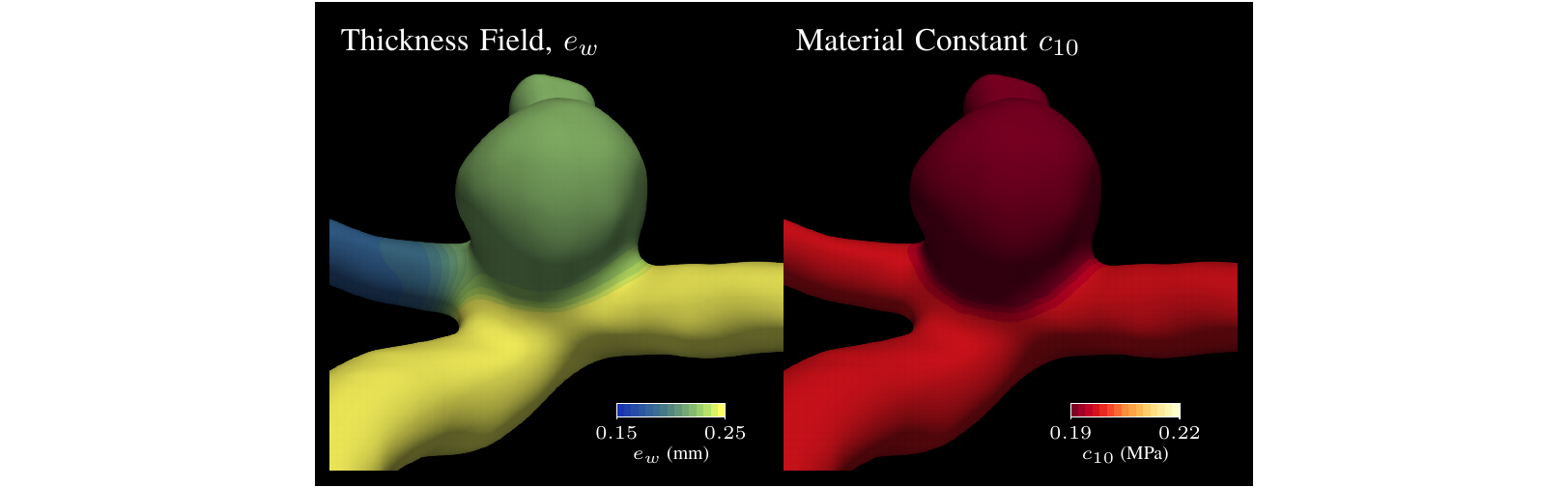}

        \caption{%
            Example of the resulting thickness and material constant
            $\gls{MooneyCoeffs}_{10}$ fields defined \emph{a priori} for
            \gls{ia} case \case{case4ruptured}.
        }

        \label{fig:wallMorphologyModeling}
\end{figure}

    Briefly, first the thickness of the branches (symbolically represented as
    \gsub{surface}{branch}) was computed based on established evidence that the
    thickness of arteries is proportional to their lumen diameter
    \citep{Fung1993}. Then, based on the thickness field of the surrounding
    branches, we estimated the patient-specific uniform thickness of the
    \gls{ia} sac surface (\gsub{surface}{aneurysm}), \gls{aneurysmThickness},
    as a weighted average of it (see the left panel of
    \cref{fig:wallMorphologyModeling}). This thickness field was then used to
    create the solid wall meshes. We assumed the material constants of the
    \gls{mr} law, \materialcoeff{MooneyCoeffs}{1}{0},
    \materialcoeff{MooneyCoeffs}{0}{1}, and \materialcoeff{MooneyCoeffs}{1}{1},
    as uniform in both the branches and aneurysm sac but with different values
    according to rupture status, following experimental evidence that
    unruptured \gls{ia} tissue is stiffer than ruptured \gls{ia} tissue
    \citep{Costalat2011}. Hence, the values for each constant (see
    \cref{tab:materialConstantsRupturedAndUnruptured}) were based on averages
    of experimental data, with ruptured and unruptured \glspl{ia}, provided by
    \citet{Costalat2011}. Regarding the material constants on
    \gsub{surface}{branch}, we used the mean values of the entire samples
    measured by \citep{Costalat2011}. This approach was preferred instead of
    using data from different studies and resulted in the material of the
    surrounding branches (i.e. healthy arteries) being less stiff than
    unruptured \glspl{ia}, as suggested by experimental investigations
    \citep{Robertson2015}.

    \begin{table}[!htb]
        \caption{%
            Material constants selected for arteries branches,
            \gsub{surface}{branch}, and the \gls{ia} sac,
            \gsub{surface}{aneurysm}, according to rupture status based on the
            experimental work of \citet{Costalat2011}.
        }
        \small
        \begin{tabular}{c S S S}
            \toprule
                \multirow{2}{*}{Constant}
            &   {\multirow{2}{*}{\gsub{surface}{branch}}}
            &   \multicolumn{2}{c}{\gsub{surface}{aneurysm}} \\
            \cmidrule(lr){3-4}
            &
            &   {Ruptured}
            &   {Unruptured} \\
            \midrule
                \materialcoeff{MooneyCoeffs}{1}{0} (\si{\mega\pascal})
            &   0.1966
            &   0.19
            &   0.19       \\
                \materialcoeff{MooneyCoeffs}{0}{1} (\si{\mega\pascal})
            &   0.0163
            &   0.026
            &   0.023      \\
               \materialcoeff{MooneyCoeffs}{1}{1} (\si{\mega\pascal})
            &   7.837
            &   1.377
            &   11.780     \\
            \bottomrule
        \end{tabular}%
        \\[\baselineskip]
        \label{tab:materialConstantsRupturedAndUnruptured}
\end{table}

    Computationally, the heterogeneous fields of \gls{wallThickness} and the
    three material constants were built with scripts in \gls{vmtk}\R and an
    in-house code based on the \gls{vtk}\R \citep{vmtk4aneurysmsBib}. See
    \cref{fig:wallMorphologyModeling} for an example of the resulting fields of
    wall thickness and the material constants
    \materialcoeff{MooneyCoeffs}{1}{0} for case \case{case4ruptured}.

            \subsection{Data Analysis} \label{sec:dataAnalysis}

    \subsubsection{Morphological Characterisation}

    By using the surface extracted from the imaging examinations, i.e.
    corresponding to the undeformed configuration, we characterised the sample
    morphologically by computing size metrics (maximum sac height, maximum sac
    diameter, aneurysm neck diameter, sac area and sac-enclosed volume) and
    shape metrics (aspect ratio, undulation, non-sphericity, and ellipticity
    indices) as defined by \citet{Ma2004,Raghavan2005}, and \citet{Dhar2008}.
    We also computed curvature-based indices proposed by \citet{Ma2004} defined
    as surface-averages and L2-norms of the Gaussian, \gls{GaussianCurvature},
    and mean, \gls{meanCurvature}, curvatures, labeled, respectively,
    \gls{areaAvgGaussianCurvature}, \gls{areaAvgMeanCurvature},
    \gls{l2NormGaussianCurvature}, and \gls{l2NormMeanCurvature} (the
    \enquote{AA} and \enquote{LN} stand for \enquote{area-average} and
    \enquote{L2-norm}, respectively).

    \subsubsection{Physical Variables of Analysis}

    From the results of the numerical simulations, we selected the main
    variables of our analysis as the largest principal Cauchy stress field of
    the solid wall, labeled \gls{maxPrincipalCauchyStress}, and the largest
    principal stretch, \gls{maxStretch}, defined as the square root of the
    largest principal value of the right Cauchy-Green deformation tensor. Both
    were taken on the luminal surface of the \emph{deformed} configuration,
    \gsupsub{surface}{lumen}{aneurysm}, at the peak-systole.

    \subsubsection{Patching of the Sac Surface}

    We performed three different \enquote{patchings} of the luminal surface of
    the \gls{ia} sac to investigate the distributions of
    \gls{maxPrincipalCauchyStress} and \gls{maxStretch}: (1) a
    abnormal-hemodynamics-based patching, (2) a \enquote{physician-oriented}
    patching, and (3) a local-shape patching based on the local curvature of
    the aneurysm sac surface.

    \subsubsection*{Abnormal-Hemodynamics Patching}

    This patching allows the investigation of connections between underlying
    hemodynamics adjacent to an aneurysm wall and the local mechanical
    response. As explained in the introduction, there are specific hemodynamic
    conditions that are associated with phenotypic and mechanical modifications
    to the wall structure of an \gls{ia}. We selected two of these
    \enquote{abnormal-hemodynamics} conditions that are associated with two
    phenotypic changes \citep{Meng2014}. \enquote{High-flow} conditions were
    defined as regions of high \gls{tawss} and low \gls{osi} --- the \gls{osi}
    is a measure of the rotation of the \gls{wss} vector in a point, defined by
    \citet{He1996} --- and were more likely to lead to the \enquote{type-I}
    wall phenotype, characterised by thin, translucent, and stiffer walls. On
    the other hand, \enquote{low-flow} conditions were defined as regions of
    flow characterised by low \gls{tawss} and high \gls{osi} and were more
    likely to cause the \enquote{type-II} phenotype, characterised by thick,
    atherosclerotic, stiffer walls.

    Specific thresholds of \gls{tawss} and \gls{osi} to identify these regions
    do not exist and have been investigated by few studies
    \citep{Furukawa2018,Cebral2019}, consequently, there is no agreement on
    their measurements. Therefore, we selected thresholds that would
    effectively create both type-I and type-II patches for the entire \gls{ia}
    sample while using values that were close to the ones found by the studies
    above. The resulting values were: $\gls{timeAvgWallShearStress} <
    \SI{5}{\pascal}$ and $\gls{oscillatoryShearIndex} > \SI{0.01}{}$ for
    low-flow hemodynamics and $\gls{timeAvgWallShearStress} > \SI{10}{\pascal}$
    and $\gls{oscillatoryShearIndex} < \SI{0.001}{}$ for high-flow
    hemodynamics.

\begin{figure}[!htb]
        \includegraphics[%
            width=\textwidth
        ]   {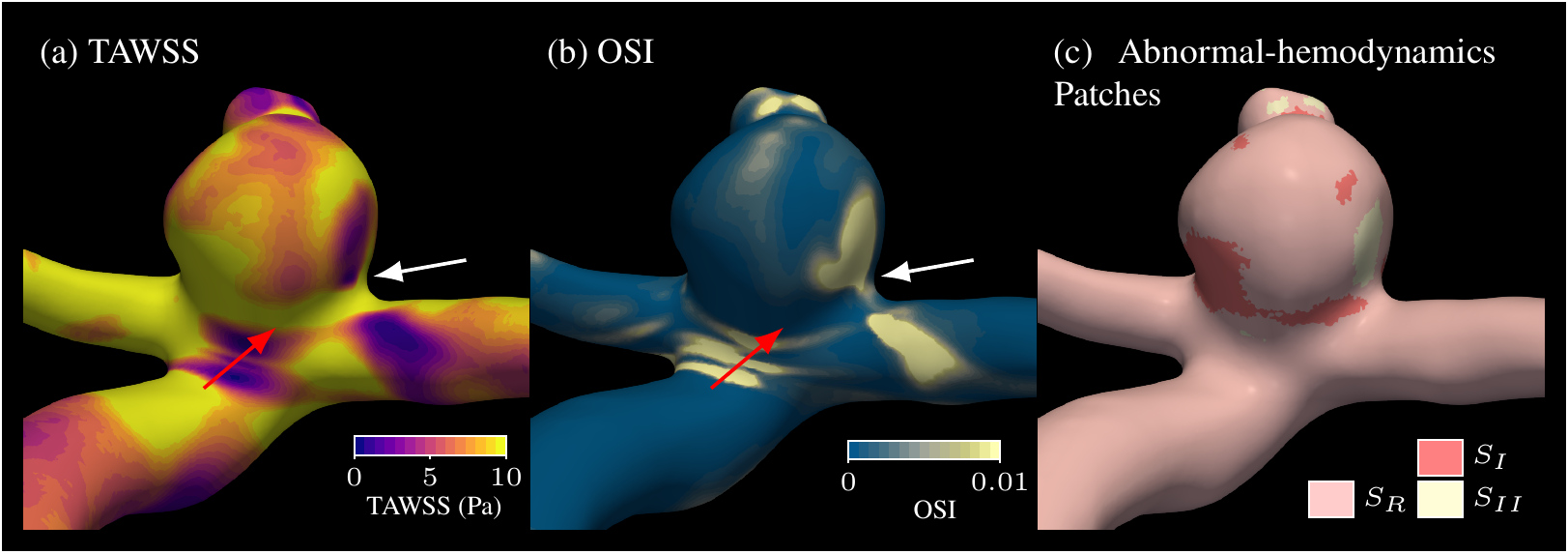}

        \caption{%
            Example of the abnormal-hemodynamics patching for \gls{ia} case
            \case{case4ruptured}: identification of high-flow (red arrows,
            type-I patches) and low-flow (white arrows, type-II patches)
            regions on the sac surface based on the (a) \gls{tawss} and (b)
            \gls{osi} fields; and (c) final patching of the sac surface;
            \enquote{regular} are patches that are neither type-I nor type-II.
        }

        \label{fig:abnormalHemodynamicsPatching}
\end{figure}

    For each patient-specific \gls{ia} geometry, these regions were identified
    after performing \gls{cfd} simulations using the rigid-wall assumption ---
    the specific methodology of these simulations can be found in
    \citet{Oliveira2021}. The \gls{tawss} and \gls{osi} were computed and the
    identification of the abnormal-hemodynamics patches was performed
    automatically by using an in-house code implemented with the \gls{vtk}\R
    library \citep{vmtk4aneurysmsBib} (see
    \cref{fig:abnormalHemodynamicsPatching}). Patches that were identified as
    neither type-I nor type-II were labelled \enquote{regular}. Finally, we
    labelled type-I, type-II, and regular patches as
    \gsub{surface}{typeIPatch}, \gsub{surface}{typeIIPatch}, and
    \gsub{surface}{normalPatch}, respectively.

    \subsubsection*{Physician-oriented Patching}

\begin{figure}[!htb]
        \caption{%
            Example for case \case{case4ruptured} of the
            \enquote{physician-oriented} patching into \enquote{neck}
            (\gsub{surface}{neckPatch}), \enquote{body}
            (\gsub{surface}{bodyPatch}), \enquote{dome}
            (\gsub{surface}{domePatch}) as defined by
            \citet{SalimiAshkezari2021}.
        }

        \includegraphics[%
            width=\textwidth
        ]   {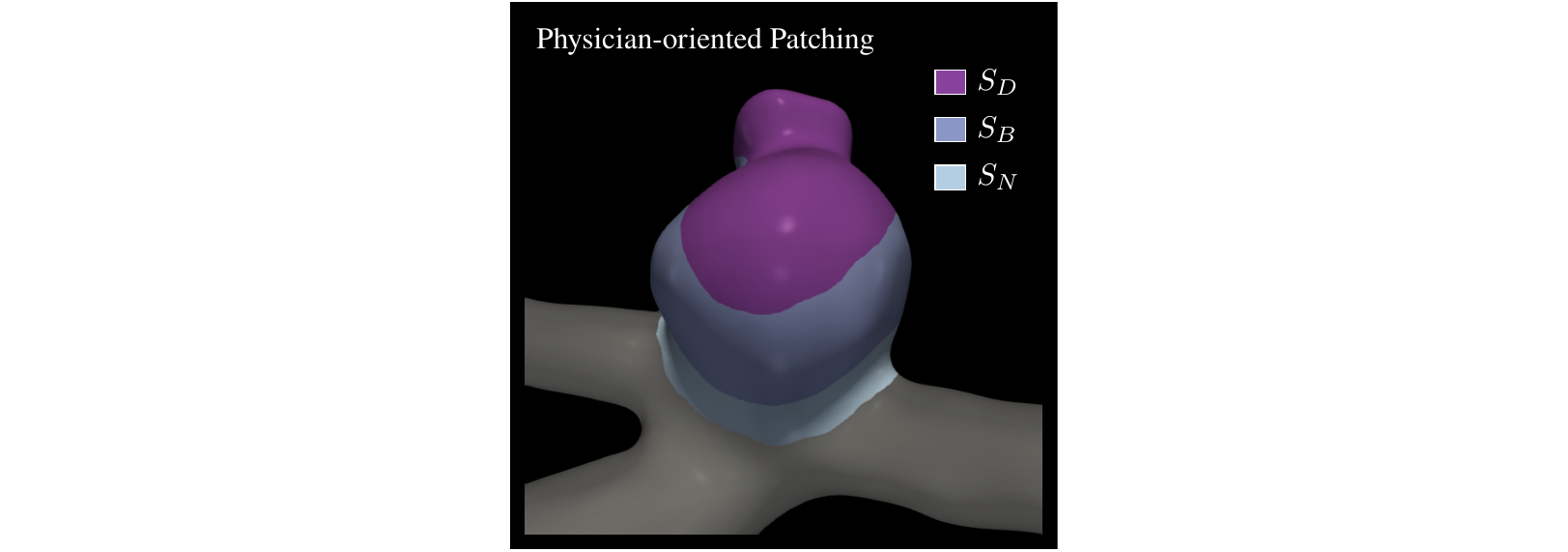}

        \label{fig:physicianOrientedPatching}
\end{figure}

    This patching was based on an classification typically employed
    by physicians to split an \gls{ia} sac into \enquote{neck}, \enquote{body},
    and \enquote{dome} regions. Although commonly employed in the medical
    practice, no formal mathematical definition is used by physicians.
    Therefore, the definition proposed by \citet{SalimiAshkezari2021} was
    employed here. To do this, the contours of the geodesic distance from the
    neck contour were computed over each \gls{ia} lumen sac surface,
    \gls{distanceToNeckLine}, using the \gls{vtk}\R library. Then, the sac was
    split into the three patches named neck, \gsub{surface}{neckPatch}, body,
    \gsub{surface}{bodyPatch}, and dome, \gsub{surface}{domePatch}, based on
    the maximum geodesic distance to the neck contour,
    $(\gls{distanceToNeckLine})_{\gls{maximum}}$, as follows: the neck was the
    region between the neck contour and \SI{20}{\percent} of
    $(\gls{distanceToNeckLine})_{\gls{maximum}}$, the body patch lied between
    \num{20} and \SI{60}{\percent} of
    $(\gls{distanceToNeckLine})_{\gls{maximum}}$ and the dome was defined as
    the region between \SI{60}{\percent} of
    $(\gls{distanceToNeckLine})_{\gls{maximum}}$ to the tip of the sac.  An
    example of the resulting patching is shown in
    \cref{fig:physicianOrientedPatching}.

    \subsubsection*{Local-shape Patching}

\begin{figure}[!htb]
        \caption{%
            Example for case \case{case4ruptured} of the (a) mean,
            \gls{meanCurvature}, and (b) Gaussian, \gls{GaussianCurvature},
            curvature fields of the lumen surface, and (c) the resulting
            local-shape patching, based on the one proposed by \citet{Ma2004}.
        }

        \includegraphics[%
            width=\textwidth
        ]   {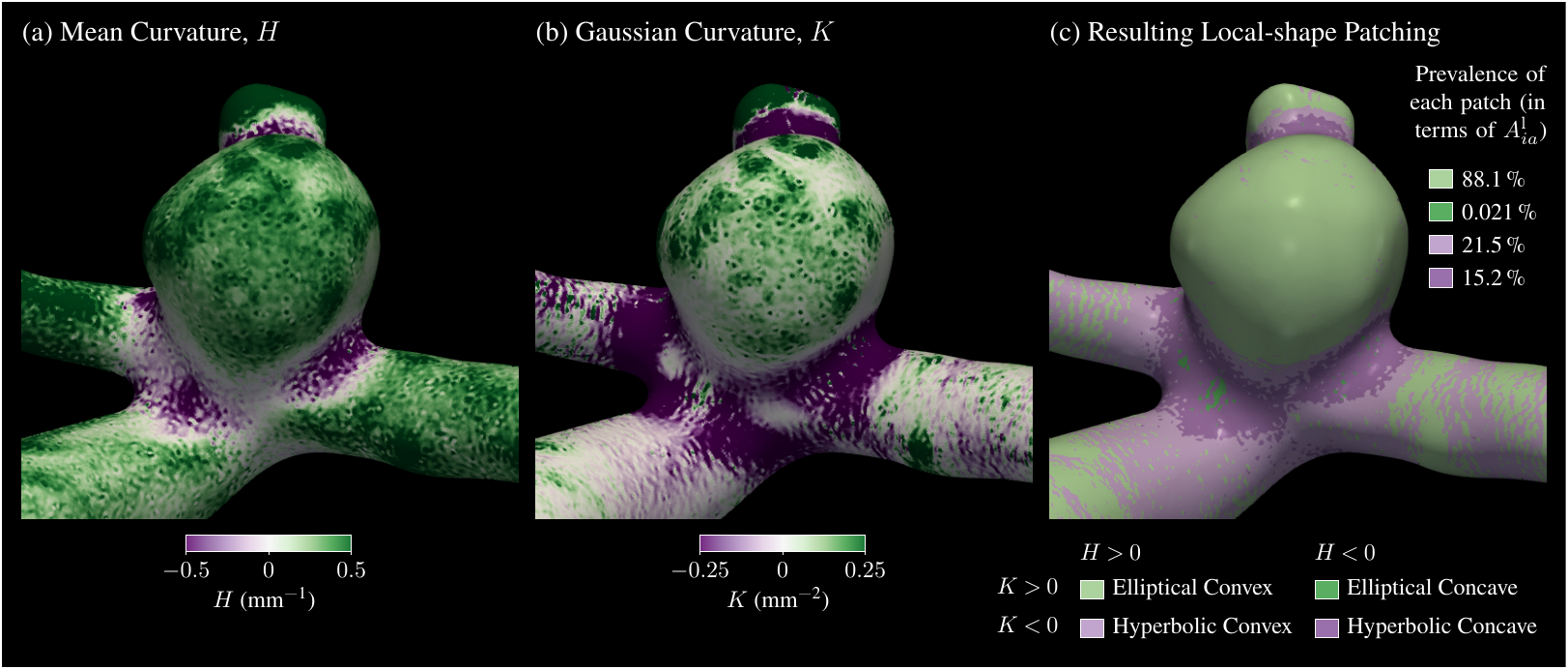}

        \label{fig:sacShapePatching}
\end{figure}

    The \enquote{local-shape} patching was based on the Gaussian,
    \gls{GaussianCurvature}, and mean, \gls{meanCurvature}, curvatures of a
    surface (see \cref{fig:sacShapePatching}a and b, respectively, illustrating
    these fields for case \case{case4ruptured}), following the classification
    presented by \citet{Ma2004}. The Gaussian curvature, given in
    \si[per-mode=reciprocal]{\per\square\milli\meter}, defines elliptic
    ($\gls{GaussianCurvature} > 0$, sphere-like), parabolic
    ($\gls{GaussianCurvature} = 0$, cylindric-like), or hyperbolic
    ($\gls{GaussianCurvature} < 0$, saddle-like) points of the surface. The
    mean curvature, given in \si[per-mode=reciprocal]{\per\milli\meter},
    defines local convexity of the surface, with $\gls{meanCurvature} > 0$
    indicating convex points and $\gls{meanCurvature} < 0$ concave points.
    These fields were calculated for the vascular surfaces using the
    \gls{vtk}\R library.

    The local-shape classification used here is shown in the table in
    \cref{fig:sacShapePatching}c --- comparing with the classification given by
    \citet{Ma2004}, we only employed the ones found on the aneurysms, although
    others are possible, such as parabolic and planar surfaces. This
    classification yielded a map of the types of regions spanning the vascular
    surface, including the aneurysms (see \cref{fig:sacShapePatching}c). For
    the subsequent analysis, elliptical and hyperbolic patches are indicated as
    \gsub{surface}{ellipticPatch} and \gsub{surface}{hyperbolicPatch},
    respectively.

    As can be seen in \cref{fig:sacShapePatching}a and b, both
    \gls{GaussianCurvature} and \gls{meanCurvature} are  \enquote{noisy}
    fields. As explained in \citet{Ma2004}, curvatures are second-order
    quantities in nature that depend on the coordinates of the surface points.
    Hence, their computation for discretised surfaces leads to noisy fields,
    and, more importantly, the surface mesh refinement level may affect the
    evaluation of the curvatures. To avoid any impact on the analysis performed
    in this work, both \gls{meanCurvature} and \gls{GaussianCurvature} were
    computed on highly-refined versions of the luminal surface and,
    subsequently, interpolated to the surfaces where the wall simulation
    results were stored. This procedure avoided discrepancies if the curvatures
    were calculated separately for each surface.

    \subsubsection{Metrics Used on the Patches}

    The prevalence of each patch type was analysed by computing their areas as
    percentages of the sac area, \gsupsub{surfaceArea}{lumen}{aneurysm}.
    Additionally, the surface average, denoted with angled brackets
    \enquote{\savg{}}, of \gls{maxPrincipalCauchyStress} and \gls{maxStretch}
    over the patch surfaces, $\gls{surface}_i$, were defined as follows, for
    \gls{maxPrincipalCauchyStress}:
    \begin{equation} \label{eq:surfaceAvgSigmaMax}
        \savg{\gls{maxPrincipalCauchyStress}}_{\gls{surface}_i}
        =
        \surfaceavg{
            \gls{surface}_i
        }{
            \gls{maxPrincipalCauchyStress}
            \functionOf{
                \gls{EulerCoord}
            }
        },
        \,\,\,\,
        \gls{EulerCoord} \in \gls{surface}_i\,,
    \end{equation}
    \noindent where $\gls{surfaceArea}\functionOf{}$ is the area operator and
    $\gls{surface}_i$ is any of the surface patches defined previously.
    Computationally, the surface average was computed using an in-house code
    that computes the integral in \cref{eq:surfaceAvgSigmaMax} with first-order
    accuracy \citep{vmtk4aneurysmsBib}.

    \subsubsection{Statistical Analysis}

    Statistical tests were performed with n = 13 and the SciPy library
    \citep{scipy} to compare both \savg{\gls{maxPrincipalCauchyStress}} and
    \savg{\gls{maxStretch}} and assuming a significance level of
    $\gls{significanceLevel} = 0.05$. We compared the metrics among the
    different patch types of the same patching, and between the ruptured and
    unruptured groups of \glspl{ia} (labeled from now on
    \enquote{\gls{rupturedGroup}} and \enquote{\gls{unrupturedGroup}},
    respectively). Normality was assessed by using the Shapiro-Wilk test. For
    the binary comparisons, the paired t-test and the Wilcoxon signed-rank
    tests for normal and non-normal distributions, respectively, were employed.
    For tertiary comparisons (for example, between the three
    abnormal-hemodynamics patchings), the ANOVA test and the Kruskal-Wallis
    test were used \emph{a priori} for normal and for non-normal distributions,
    respectively.  Subsequently, pair-wise post hoc analyses were performed to
    test the distributions. The t-test and Dunn's (with the Bonferroni
    p-adjustment) posthoc methods were employed, in this case, for normal and
    non-normal distributions, respectively, via Python's scikit-posthoc
    library.

        \section{Results} \label{sec:results}

    \subsection*{Morphological Characterisation}

    \cref{tab:meanIndices} shows the mean and \gls{sd} of the size and shape
    metrics and the curvature-based indices per rupture-status group. The
    unruptured cases are comparatively larger than the ruptured ones in terms
    of both \gls{1d} and \gls{3d} size metrics; still, this difference was not
    statistically significant (see p-values in the table). The difference
    between ruptured and unruptured cases was statistically significant only
    for the undulation, non-sphericity, and ellipticity indices and also the
    \gls{l2NormMeanCurvature}. The difference between groups
    \gls{rupturedGroup} and \gls{unrupturedGroup} for the other metrics was not
    statistically significant, including the aspect ratio, which has previously
    been used to categorise ruptured and unruptured aneurysms \citep{Weir2003}.

    \begin{table}[!htb]
        \caption{%
            Means and \gls{sd} of \gls{1d}, \gls{2d}, and \gls{3d} size and
            shape metrics and curvature-based indices for the ruptured
            (\gls{rupturedGroup}) and unruptured (\gls{unrupturedGroup})
            subgroups of the \gls{ia} sample. Size metrics are the maximum neck
            diameter, \gls{neckDiameter}, the maximum dome diameter,
            \gls{domeDiameter}, the maximum normal height, \gls{domeHeight},
            and the sac's surface area and enclosed volume,
            \gsub{surfaceArea}{aneurysm} and \gsub{volume}{aneurysm},
            respectively. Shape indices are the aspect ratio,
            \gls{aspectRatio}, the undulation index, \gls{undulationIndex}, the
            non-sphericity index, \gls{nonsphericityIndex}, and the ellipticity
            index, \gls{ellipticityIndex}, and the curvature-based ones were
            proposed by \citep{Ma2004}.
        }
        \sisetup{%
            %table-parse-only, % disable alignment of decimal digit
            per-mode=reciprocal,
            table-format=3.3(2)
        }
        \small
        \begin{tabular}{
            c
            c
            S[table-align-uncertainty=true]
            S[table-align-uncertainty=true]
            S[table-format=1.3,table-comparator=true]
        }
            \toprule
                Metric Type
            &   Metric
            &   \gls{rupturedGroup}
            &   \gls{unrupturedGroup}
            &   \emph{p-value} \\
            \midrule
                \multirow{2}{*}{%
                    \parbox{3cm}{%
                        \centering
                        \gls{1d} Size Indices
                    }
                }
            &   {\gls{domeDiameter} (\si{\milli\meter})}
            &   4.93 \pm 1.59 & 6.28 \pm 1.30 & 0.12 \\
            &   {\gls{domeHeight} (\si{\milli\meter})}
            &   5.14 \pm 1.68 & 5.52 \pm 1.40  & 0.67 \\
            \cmidrule(lr){1-5}
                \multirow{2}{*}{%
                    \parbox{3cm}{%
                        \centering
                        \gls{3d} Size Indices
                    }
                }
            &   {\gsub{surfaceArea}{aneurysm} (\si{\square\milli\meter})}
            &   91.04 \pm 57.23 & 115.74 \pm 51.65 & 0.43 \\
            &   {\gsub{volume}{aneurysm} (\si{\square\milli\meter})}
            &   90.18 \pm 78.89 & 140.64 \pm 87.52 & 0.37 \\
            \cmidrule(lr){1-5}
                \parbox{3cm}{%
                    \centering
                    \gls{2d} Shape Index
                }
            &   {\gls{aspectRatio}}
            &   1.34 \pm 0.34 & 1.13 \pm 0.17 & 0.18 \\
            \cmidrule(lr){1-5}
                \multirow{3}{*}{%
                    \parbox{2cm}{%
                        \centering
                        \gls{3d} Shape Indices
                    }
                }
            &   \gls{undulationIndex}
            &   0.14 \pm 0.024 & 0.077 \pm 0.015 & < 0.001 \\
            &   \gls{ellipticityIndex}
            &   0.14 \pm 0.042 & 0.091 \pm 0.030 & 0.025 \\
            &   \gls{nonsphericityIndex}
            &   0.20 \pm 0.028 & 0.13 \pm 0.025 & < 0.001\\
            \cmidrule(lr){1-5}
                \multirow{4}{*}{%
                    \parbox{2cm}{%
                        \centering
                        Curvature-based Indices
                    }
                }
            &   \gls{areaAvgMeanCurvature} (\si{\per\milli\meter})
            &   0.33 \pm 0.09 & 0.27 \pm 0.057 & 0.14 \\
            &   \gls{areaAvgGaussianCurvature} (\si{\per\square\milli\meter})
            &   0.059 \pm 0.037 & 0.052 \pm 0.029 & 0.89 \\
            &   \gls{l2NormGaussianCurvature}
            &   2.98 \pm 1.07 & 2.08 \pm 0.678 & 0.091 \\
            &   \gls{l2NormMeanCurvature}
            &   0.34 \pm 0.038 & 0.29 \pm 0.025 & 0.020 \\
            \bottomrule
        \end{tabular}%
        \\[\baselineskip]
        \label{tab:meanIndices}
    \end{table}

    \subsection*{Patch Type Prevalence Across the Sample}

\begin{figure}[!htb]
        \caption{%
            Distributions of the ratio between patch area to the aneurysm sac
            surface area, \gsupsub{surfaceArea}{lumen}{aneurysm}, of the (a)
            abnormal-hemodynamics patches of type I, II, and \enquote{regular}
            and (b) of the local-shape types, grouped by convexity, for the
            \gls{ia} sample.
        }

        \includegraphics[%
            width=\textwidth
        ]   {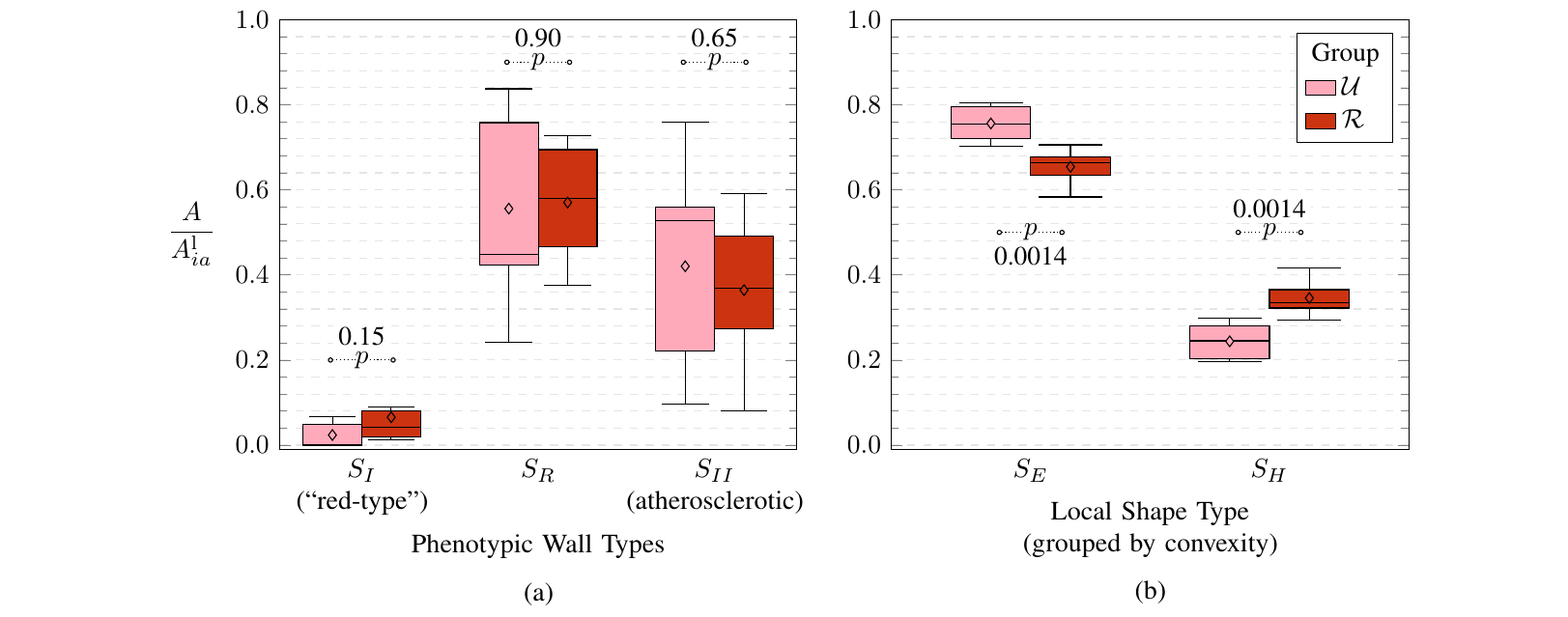}

        \label{fig:iaPatchTypesPrevalence}
    \end{figure}

    The classification defined in this work to identify the type-I and type-II
    patches led to a prevalence of each as shown in
    \cref{fig:iaPatchTypesPrevalence}a. Type-II patches (atherosclerotic) tend
    to be significantly more prevalent than type-I ones ($\gls{pValue} <
    \num{0.001}$), with some unruptured cases presenting type-II patches only.
    Nevertheless, given that the samples taken include \glspl{ia} spanning an
    extensive range of sizes in different sites, implying different
    hemodynamics, it is striking how rare type-I patches seem to be.
    Furthermore, note that the described prevalence occurred independently of
    rupture status, with no statistically significant differences between
    groups \gls{rupturedGroup} and \gls{unrupturedGroup}, as can bee seen by
    inspecting the p-values in \cref{fig:iaPatchTypesPrevalence}a. In this
    work, p-values were depicted in the plots themselves with the symbol
    \enquote{\tikz{\pic {picStatsTest=north/{}};}}, where the circles indicate
    the distributions being compared.

    Regarding the local-shape patching, hyperbolic patches were less prevalent
    in this sample. They occupied a range of \SIrange{30}{40}{\percent} of the
    \glspl{ia} sac area, as can be seen in \cref{fig:iaPatchTypesPrevalence}b.
    In the figure, note that both convex and concave patches were added
    together and included under the elliptic and hyperbolic points. This was
    performed because elliptical concave regions accounted for less than
    \SI{0.1}{\percent} of the sac area, on average. Furthermore, hyperbolic
    patches were significantly larger in ruptured \glspl{ia} than unruptured
    ones (mean $\pm$ \gls{sd} $=$ \SI{34.6 \pm 4.3}{\percent} \emph{versus}
    \SI{24.4 \pm 4.3}{\percent}, respectively; see p-values in
    \cref{fig:iaPatchTypesPrevalence}b).

    \subsection*{%
        Local Mechanical Response on Different Wall Patches
    }   \label{sec:localVarOnPatches}

\begin{figure}[!t]
        \caption{%
            Distributions of (a) \savg{\gls{maxPrincipalCauchyStress}} and (b)
            \savg{\gls{maxStretch}} over the surface patches of
            elliptic and hyperbolic types, segregated by rupture-status groups.
        }

        \includegraphics[%
            width=\textwidth
        ]   {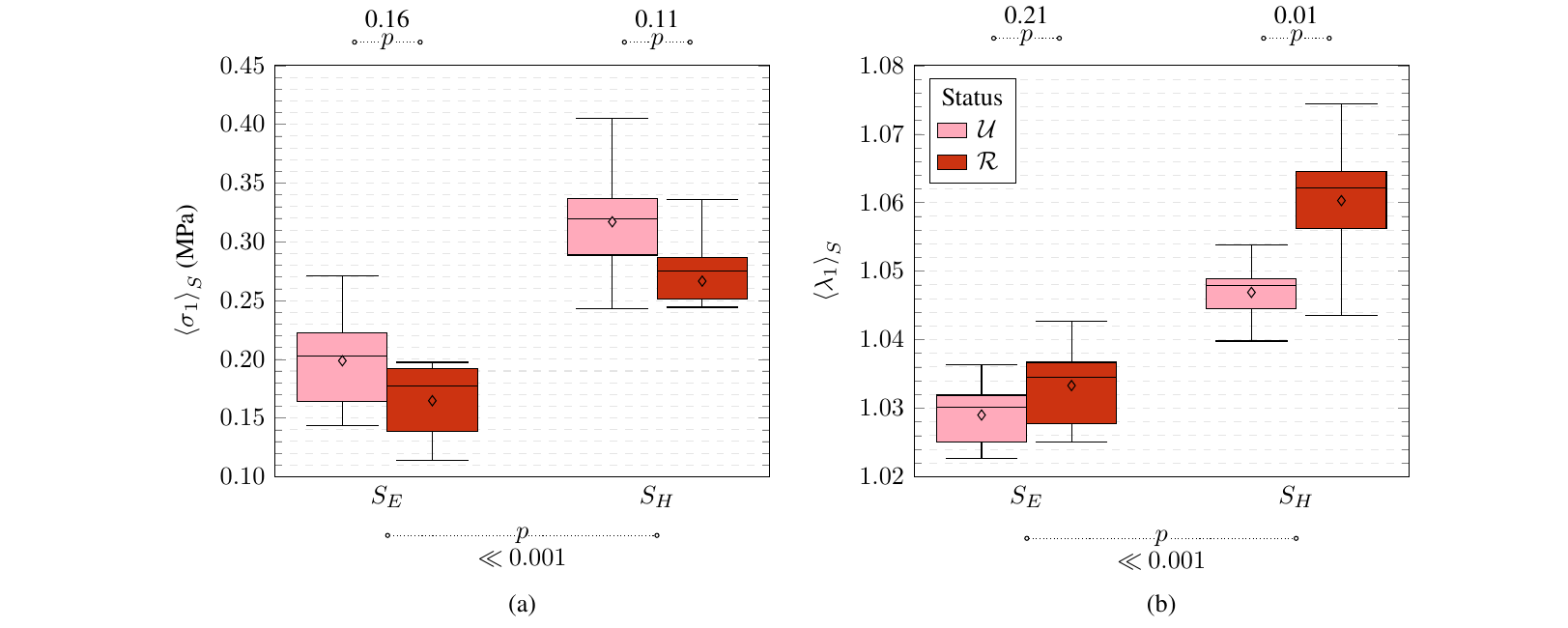}

        \label{fig:boxPlotsPerLocalShapeRuptVsUnrupt}
    \end{figure}

    Hyperbolic patches display significantly higher
    \savg{\gls{maxPrincipalCauchyStress}} --- as high as twice in some cases
    --- compared to elliptic ones when comparing the whole sample, and no
    significant differences were found between the \gls{rupturedGroup} and
    \gls{unrupturedGroup} sub-groups (see
    \cref{fig:boxPlotsPerLocalShapeRuptVsUnrupt}a and the p-values in it).
    Significant differences were also found for \savg{\gls{maxStretch}} between
    elliptic and hyperbolic patches (see
    \cref{fig:boxPlotsPerLocalShapeRuptVsUnrupt}b). Furthermore, the largest
    levels of \savg{\gls{maxStretch}} were found on hyperbolic patches of the
    ruptured group, with statistical significance. In contrast, no significant
    differences between groups \gls{rupturedGroup} and \gls{unrupturedGroup}
    were found for the elliptic ones (see p-values in
    \cref{fig:boxPlotsPerLocalShapeRuptVsUnrupt}b).

\begin{figure}[!htb]
        \caption{%
            Distributions of \savg{\gls{maxPrincipalCauchyStress}} (a and b),
            \savg{\gls{maxStretch}} (c and d), and
            \savg{\gls{GaussianCurvature}} (e and f) over the
            abnormal-hemodynamics patches (\gsub{surface}{typeIPatch},
            \gsub{surface}{typeIIPatch}, and \gsub{surface}{normalPatch}; left
            column) and over the physician-oriented patches
            (\gsub{surface}{neckPatch}, \gsub{surface}{bodyPatch}, and
            \gsub{surface}{domePatch}; right column), segregated by
            rupture-status groups.
        }

        \includegraphics[%
            width=\textwidth
        ]   {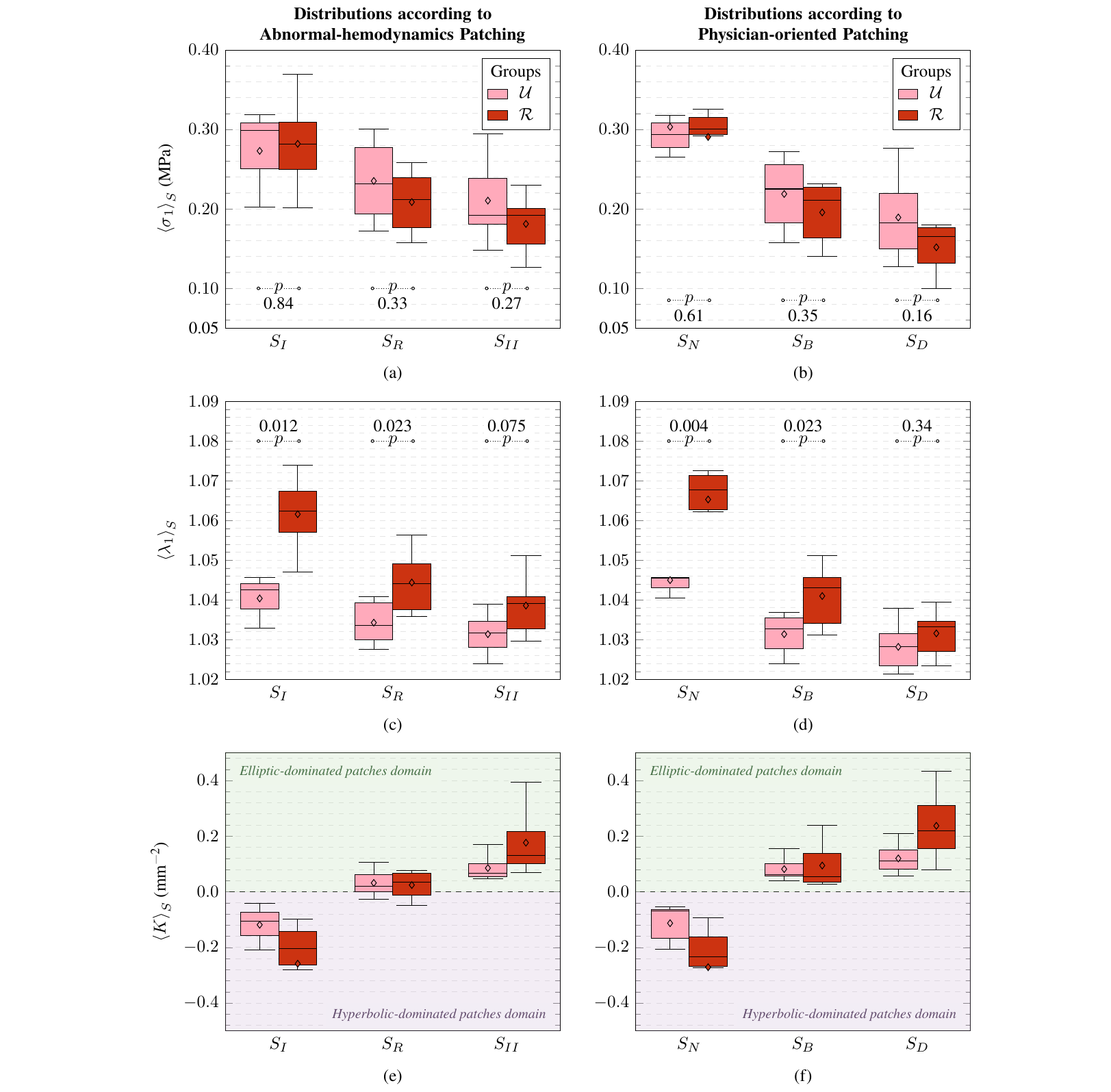}

        \label{fig:boxPlotsPerLocalPatchesRuptVsUnrupt}
    \end{figure}

    Both \savg{\gls{maxPrincipalCauchyStress}} and \savg{\gls{maxStretch}} were
    higher on type-I patches by as much as twice compared to type-II ones. At
    the same time, intermediate levels of those variables were found on the
    regular patches (see \cref{fig:boxPlotsPerLocalPatchesRuptVsUnrupt}a and
    c). Statistical tests confirmed that there were significant differences
    between the three patch types for both
    \savg{\gls{maxPrincipalCauchyStress}} and \savg{\gls{maxStretch}}
    ($\gls{pValue} = \num{0.0019}$ and $\gls{pValue} < \num{0.001}$,
    respectively), although a post hoc pair-wise analysis showed that
    \savg{\gls{maxPrincipalCauchyStress}} was only significantly different
    between the type-I and type-II patches, whereas \savg{\gls{maxStretch}} was
    significantly different between the type-I and both type-II and regular
    patches. It is important to highlight that only the cases with non-zero
    type-I patch areas were included in this comparison.

    Similarly, the neck patch exhibited significantly higher levels of both
    \savg{\gls{maxPrincipalCauchyStress}} and \savg{\gls{maxStretch}} compared
    to the body and dome surface-averages ($\gls{pValue} < \num{0.001}$ for
    both variables by comparing the three distributions). The post hoc
    pair-wise analysis showed no significant differences for both variables
    between the body and dome patches.

    Irrespective of the abnormal-hemodynamics or physician-oriented patchings,
    \savg{\gls{maxPrincipalCauchyStress}} was not significantly different
    between the ruptured and unruptured groups (see p-values in
    \cref{fig:boxPlotsPerLocalPatchesRuptVsUnrupt}a and
    \ref{fig:boxPlotsPerLocalPatchesRuptVsUnrupt}b). On the other hand,
    \savg{\gls{maxStretch}} was found to be significantly higher in ruptured
    cases than in unruptured ones only on type-I and regular patches, in the
    abnormal-hemodynamics patching (see p-values in
    \cref{fig:boxPlotsPerLocalPatchesRuptVsUnrupt}c). The same was found on the
    neck and body patches, in the physician-oriented patching (see
    \ref{fig:boxPlotsPerLocalPatchesRuptVsUnrupt}d). Note, however, that the
    most significant differences occurred on type-I and neck patches.

    The evident similarity described above is not a coincidence; it is related
    to the underlying curvature of the abnormal-hemodynamics and
    physician-oriented patches. As can be seen in
    \cref{fig:boxPlotsPerLocalPatchesRuptVsUnrupt}e and
    \ref{fig:boxPlotsPerLocalPatchesRuptVsUnrupt}f, which show the
    distributions of \savg{\gls{GaussianCurvature}} for the
    abnormal-hemodynamics and physician-oriented patchings (while suitably
    divided into elliptic and hyperbolic regions of $\gls{GaussianCurvature} >
    0$ and $\gls{GaussianCurvature} < 0$, respectively),
    \savg{\gls{GaussianCurvature}} on type-I and neck patches are, on average,
    entirely in the negative-Gaussian region, showing that they contain
    predominantly hyperbolic points. On the other hand,
    \savg{\gls{GaussianCurvature}} on type-II and dome patches contain more
    elliptic patches. Intermediate situations occur for the regular and body
    patches. Therefore, the largest \savg{\gls{maxPrincipalCauchyStress}} and
    \savg{\gls{maxStretch}} seen on the neck and type-I patches are explained
    by their curvature (see \cref{fig:boxPlotsPerLocalShapeRuptVsUnrupt}).

    Further support for this relationship between the mechanical variables and
    the curvature was found by plotting the
    \savg{\gls{maxPrincipalCauchyStress}} and \savg{\gls{maxStretch}} against
    \savg{\gls{GaussianCurvature}} and \savg{\gls{meanCurvature}} over elliptic
    and hyperbolic patches (see \cref{fig:corrMechanicalVarVsCurvatures} in the
    supplementary material). We found significant negative correlations between
    the mechanical variables and both curvatures (see Pearson's correlation
    coefficient, \gls{PearsonCoeff}, on top of each plot in
    \cref{fig:corrMechanicalVarVsCurvatures}). Although a continuous
    relationship between the stress and stretch and the wall curvature is
    probably harder to find, our results suggest that such a relationship
    may exist for the wall of \glspl{ia} and is most likely to be similar to
    Laplace's law, i.e. both stress and stretch are inversely proportional to
    the curvatures. Furthermore, and most importantly, this regional analysis
    suggests that one of the most critical determinants of the largest
    \gls{maxPrincipalCauchyStress} and \gls{maxStretch} in \glspl{ia} walls is
    the wall curvature.

        \section{Discussions} \label{sec:discussion}

    \subsection*{%
        The Connections Between Mechanical Response, Hemodynamics, and Rupture
    }

    The connection between curvature and the mechanical response of the
    \gls{ia} walls shown by these results is not a complete surprise. However,
    our work has helped identify exactly which curvature metric is important.
    An \gls{ia} wall, although with a complex tissue composition, can be
    assumed, structurally, as a thick shell, hence its curvature is an
    intrinsic property, and it is known to play an essential role in
    determining the mechanical response of such structures. \citet{Ma2007}
    investigated the effect of curvature on \glspl{ia} wall mechanics, although
    with numerical results based on simulations with static \glspl{bc} --- a
    uniform pressure applied on the luminal surface --- and using as geometry
    only the \gls{ia} sac, i.e. without the branches walls. They employed an
    anisotropic Fung-like material law and found no statistically significant
    differences for stress and strain between ruptured and unruptured
    \glspl{ia}. However, they show that their levels in the ruptured group were
    generally higher, which agrees with the findings presented in this work for
    the stretch metrics, although we have found significant differences. More
    importantly, the authors also investigated the maximum and minimum
    curvature fields and found strong correlations between stress and strain
    and the curvature.

    Our findings strongly indicate that the regions with dangerously-high
    levels of stretch and stress can be identified by looking at the local wall
    curvature. More specifically, by the regions identified as hyperbolic, i.e.
    with negative Gaussian curvature. Moreover, in those patches, the largest
    stretch was significantly higher in ruptured \glspl{ia} than in unruptured
    ones. Two reasons explain this last observation. First, due to the less
    stiff properties of the ruptured cases used in our modelling, which tends
    to produce higher stretches comparatively with unruptured ones. Second,
    the ruptured \glspl{ia} have more lobular regions and daughter sacs  or
    \enquote{blebs} (see \cref{tab:meanIndices} where the higher \gls{3d} shape
    indices of ruptured \glspl{ia} indicate this). The morphological formations
    increase the area of hyperbolic patches locally, as can be verified by the
    higher prevalence of hyperbolic patches in the ruptured group (see
    \cref{fig:iaPatchTypesPrevalence}b), which hence further increases the
    stretch levels overall on the sac.

    The wall curvature also explains the corresponding high levels of
    \gls{maxPrincipalCauchyStress} and \gls{maxStretch} on the
    abnormal-hemodynamics and physician-oriented patchings. Clinical
    observations show that \glspl{ia} invariably rupture at their dome,
    although there is a small incidence of rupture sites closer to their neck
    \citep{Park2012}. As shown by our results, the dome patch of an \gls{ia} is
    dominated by elliptic points and most likely to be composed of type-II
    patches, i.e. with an atherosclerotic phenotype (see
    \cref{fig:boxPlotsPerLocalPatchesRuptVsUnrupt}e and
    \ref{fig:boxPlotsPerLocalPatchesRuptVsUnrupt}f). By their curvature, these
    patches are more likely to have relatively low \gls{maxStretch} and
    \gls{maxPrincipalCauchyStress} levels and, moreover, due to their
    underlying hemodynamics, they are more likely to be thicker and stiffer,
    further decreasing the overall values of these variables. Therefore, other
    factors may play a role in explaining the high incidence of ruptures on the
    dome, either related to mechanical or failure properties of the wall. For
    example, it could be that mechanical variables other than stress and
    stretch play a role in causing these ruptures. Alternatively, the ultimate
    stress and ultimate strain at the dome could be smaller than the rest of
    the sac, although we have not found any research suggesting that.
    Nevertheless, a realistic possibility that would explain this high
    incidence of dome ruptures is calcification, which is highly prevalent in
    human \glspl{ia} as recent findings suggest \citep{Gade2019}, especially in
    atherosclerotic regions of their wall. Further studies show that
    calcification regions in \glspl{ia} walls concentrate stress and
    significantly influence the mechanics there \citep{Fortunato2021}.
    Unfortunately, to assess this hypothesis, mechanical and imaging analysis
    of the walls of the \glspl{ia} used within this work would be necessary to
    determine the presence of calcified zones, but they were not available.

    Although the incidence of rupture sites on the neck is smaller than on the
    dome, the mechanical data presented here satisfactorily explains the
    rupture in these sites based on the local mechanical response. Both
    \gls{maxStretch} and \gls{maxPrincipalCauchyStress} were comparatively
    higher on the neck, which was also shown to be dominated by type-I patches,
    i.e. \enquote{red-patches} that are thinner and stiffer, thus slightly
    increasing both \gls{maxPrincipalCauchyStress} and \gls{maxStretch}.
    Nonetheless, failure properties should also be assessed to confirm the
    likelihood of rupture. \citet{Cebral2015a} investigated failure mechanics
    and the hemodynamics of a sample of eight unruptured \glspl{ia} and found
    that the tissue ultimate strain is negatively correlated with time-averaged
    and surface-averaged \gls{wss} on the sac lumen. Although they did not
    investigate the local properties of each \gls{ia} wall, that trend suggests
    that, as areas of high \gls{wss} get larger in an \gls{ia} lumen, its
    ultimate wall strain gets smaller. Moreover, with larger areas of high
    \gls{wss}, it is also more likely that type-I patches will exist (defined
    by high \gls{tawss} and low \gls{osi}), which in turn tend to exist under
    hyperbolic points of the lumen surface, inducing higher stresses and
    stretches. With a small ultimate strain, this situation could reasonably
    explain a neck rupture.

\begin{figure}[!htb]
        \caption{%
            Examples of (a) an unruptured \gls{ia}, case \case{case64}, with a
            thrombus or bleb near the neck patch that introduces new local
            hyperbolic patches on the neck (and, in this case, on the body
            patch too); and (b) a ruptured \gls{ia}, case \case{case4ruptured},
            harbouring a prominent bleb on its dome patch, that also creates
            new local hyperbolic patches (the white line marks the neck
            contour, and the yellow lines mark the contours between the
            \gsub{surface}{neckPatch}, \gsub{surface}{bodyPatch}, and the
            \gsub{surface}{domePatch} patches).
        }

        \includegraphics[%
            width=\textwidth
        ]   {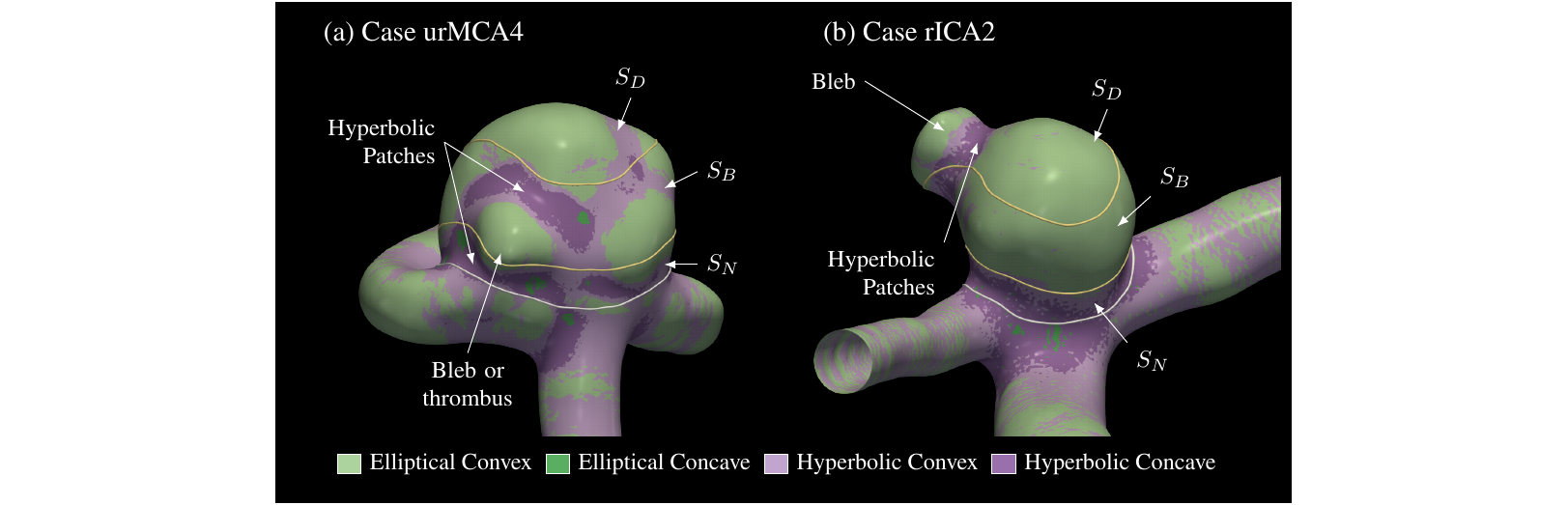}

        \label{fig:examplesPathwaysToIaRupture}
    \end{figure}

    Moreover, according to the pathways to rupture proposed by
    \citet{Meng2014}, red-type and atherosclerotic regions may eventually lead
    to blebs and thromboses on an \gls{ia} wall, respectively. These focal
    manifestations effectively create local hyperbolic regions that concentrate
    stress and stretch in both cases (see the two examples shown in
    \cref{fig:examplesPathwaysToIaRupture} with \gls{ia} cases of the sample
    used in this work). Hence, further increasing the likelihood of rupture in
    those situations. Nevertheless, it is important to reinforce that the two
    mechanisms described in the previous paragraphs depend on particularities
    of \gls{ia} failure properties, which should be investigated in future
    studies.

    Interestingly, too, is the relationship between local shape and the
    abnormal-hemodynamics patches. The data revealed that, in the sample used
    here, type-I and type-II patches have larger areas of hyperbolic and
    elliptic patches, respectively. Because both type-I and type-II patches
    were defined based on particular hemodynamic environments, the local lumen
    curvature could also determine local hemodynamics. From the fluid mechanics
    perspective, this is likely because the curvature of a surface is an
    important geometric feature that determines the dynamics of the boundary
    layer adjacent to it. But whether local curvature causes the low-flow and
    high-flow hemodynamics is harder to assess due to the temporal and spatial
    complexity of the flow inside an \gls{ia}. We found no study on
    cardiovascular fluid mechanics that investigated this relationship.

\begin{figure}[!htb]
        \caption{%
            Distributions of (a) \savg{\gls{timeAvgWallShearStress}} and (b)
            \savg{\gls{oscillatoryShearIndex}} over the surface patches of
            elliptic and hyperbolic types, for the \gls{ia} sample, segregated
            by rupture-status groups.
        }

        \includegraphics[%
            width=\textwidth
        ]   {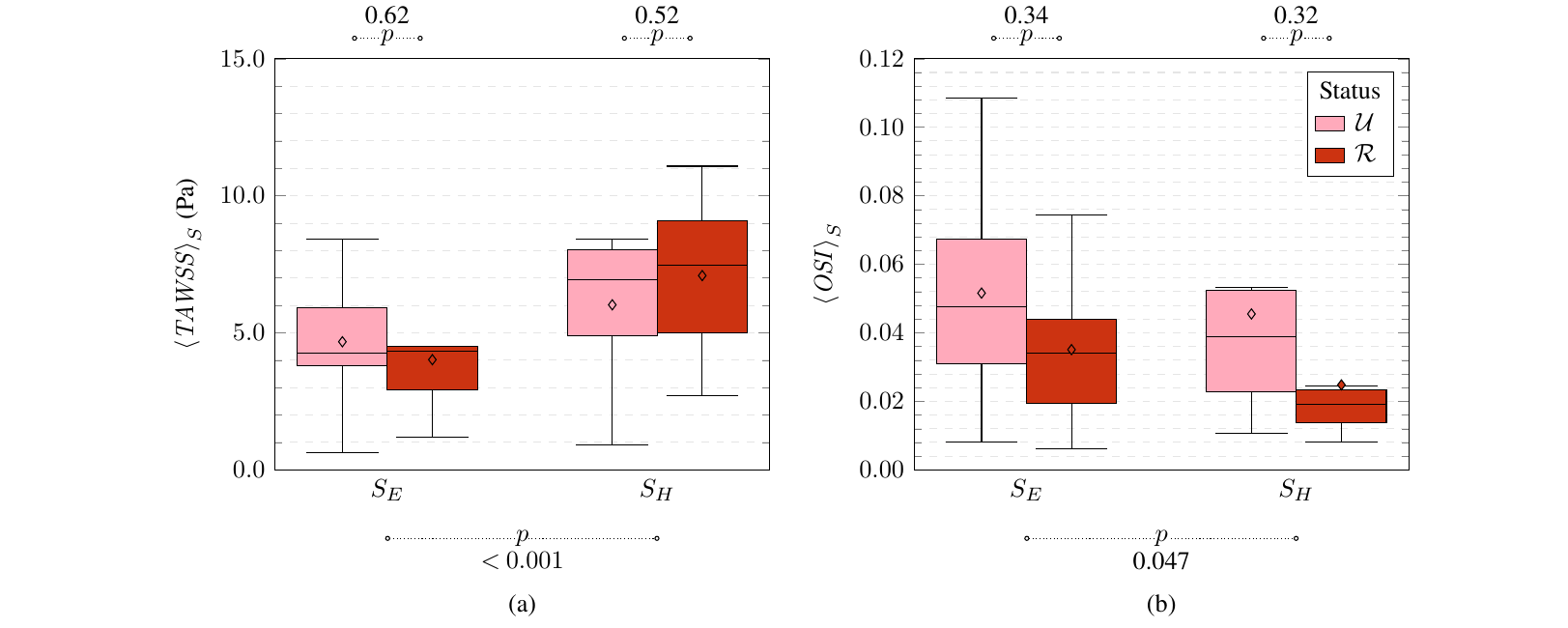}

        \label{fig:boxPlotsLocalHemodynamics}
    \end{figure}

    It is important to note that the prevalence of type-I and type-II patches
    may depend on the particular thresholds we have chosen. Nonetheless, even
    if we have changed those values slightly (while still close to the values
    found by the studies on the subject), it is unlikely that that prevalence
    would change substantially. To assess this possible relationship, at least
    for the samples used in this work and independently of the threshold values
    used to build the abnormal-hemodynamics patching, we performed an analysis
    of the hemodynamics data by computing the surface-average of \gls{tawss}
    and \gls{osi} on elliptical and hyperbolic patches (see
    \cref{fig:boxPlotsLocalHemodynamics}). Indeed, the \gls{tawss} was
    significantly higher on hyperbolic patches than on elliptic ones, whereas
    the contrary was the case for the \gls{osi}. Moreover, no significant
    differences were found for either \savg{\gls{timeAvgWallShearStress}} or
    \savg{\gls{oscillatoryShearIndex}} between ruptured and unruptured groups.
    This explains the higher prevalence of type-I patches on hyperbolic
    patches. However, it is important to insist that a causal explanation for
    this relationship must be the subject of future investigations.

\begin{figure}[!htb]
        \caption{%
            Pathways that could potentially explain the rupture of \glspl{ia}
            based on the mechanical response on (a) hyperbolic patches and (b)
            elliptic patches. The geometrical and biological connections
            (based on proposals by \citet{Meng2014}) between the different
            regions and phenotypes of the \gls{ia} wall are also depicted.
        }

        \includegraphics[%
            width=\textwidth
        ]   {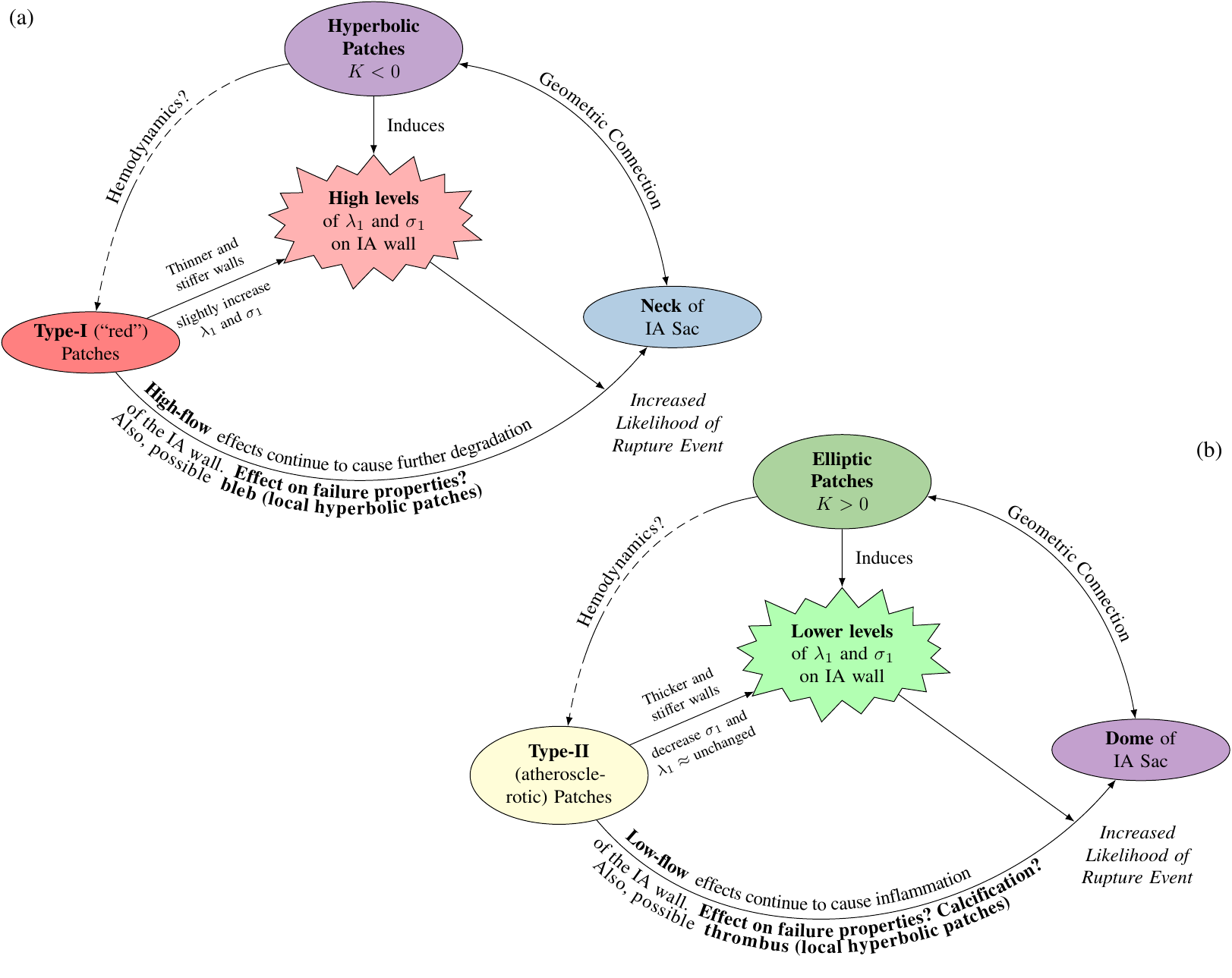}

        \label{fig:diagramMechanicalPathwaysToIaRupture}
    \end{figure}

    The connections between the mechanical response of \gls{ia} walls, their
    shape and geometry, hemodynamics, biomechanical and failure properties are
    described and exemplified in the previous paragraphs, and an eventual
    rupture event graphically summarised in the diagrams shown in
    \cref{fig:diagramMechanicalPathwaysToIaRupture}. They also embed the
    mechanobiological pathways proposed by \citet{Meng2014}, based on the
    abnormal-hemodynamics patching used in this work. To the authors'
    knowledge, this is the first attempt to categorise \gls{ia} rupture events
    based on the actual mechanical response of an \gls{ia} sample. It attempts
    to make the role that curvature plays in the mechanics of this disease
    clearer. It is important to mention that the mechanisms summarised in
    \cref{fig:diagramMechanicalPathwaysToIaRupture} must not be regarded
    statically, i.e. for an instant in time. On the contrary, due to the
    evolving nature of an \gls{ia}, these mechanisms constantly act in
    different regions of the sac wall. This certainly helps to explain the
    intriguing fact about \glspl{ia} that they can rupture while still small,
    with some reports showing that approximately \SI{50}{\percent} of the
    ruptured cases are of small-sized aneurysms \citep{Zheng2019}. Hyperbolic
    patches may occur at any time during the development of an \gls{ia},
    independently of its size. Indeed, note that hyperbolic patches are
    dominant on the neck of the \gls{ia} and, rigorously, the neck is the first
    indication of \gls{ia} inception. Therefore since early in an \gls{ia}
    development, the local level of stretch and stress may already be high on
    the neck compared to its surrounding arteries.

    \subsection*{Curvature-based Metrics to Indicate Rupture}

    The primary role played by the curvature in locating the patches with
    higher stress and stretch brings a crucial practical outcome. The wall
    curvature can be readily calculated from medical imaging examinations
    routinely performed in the diagnostics of \glspl{ia}. Therefore, physicians
    should be aware of this important relationship and consider it in their
    prognostics, although it is still combined with other clinical information
    they already use. Currently, this is somewhat accounted for based on the
    known observation that aneurysms with lobular regions or blebs are more
    likely to rupture \citep{Lindgren2016, Abboud2017}. But, more importantly,
    this finding also points to identifying curvature-based \gls{ia} sac
    metrics to help predict the rupture.

    Few attempts have been made in this direction. Curvature-based metrics were
    introduced in the context of saccular \glspl{ia}, primarily by
    \citet{Ma2004}, who proposed four curvature-based indices that give an
    overall measure of the curvature of an \gls{ia} sac. We calculated them for
    the sample used here (see \cref{tab:meanIndices}). As pointed out by
    \citet{Ma2004}, the \gls{areaAvgMeanCurvature} and
    \gls{areaAvgGaussianCurvature} depend on both size and shape, whereas
    \gls{l2NormMeanCurvature} and \gls{l2NormGaussianCurvature} rely only on
    the shape and, additionally, these indices are easier to interpret only in
    a relative basis. For example, the \gls{l2NormGaussianCurvature} was
    defined so that it is equal to the unity for a sphere; hence the means in
    \cref{tab:meanIndices} that the unruptured \glspl{ia} are more similar to a
    sphere compared to the ruptured ones. Alternatively, the ruptured
    \glspl{ia} have more saddle patches, as already suggested by inspecting
    other shape metrics (see the \gls{3d} shape indices in
    \cref{tab:meanIndices}). When comparing the metrics, we found similar
    results reported by \citet{Raghavan2005}, who assessed the potential to
    discriminate ruptured \glspl{ia} by size, shape, and curvature-based
    indices (with the \gls{l2NormMeanCurvature} and
    \gls{l2NormGaussianCurvature} among them). The authors found that the
    \gls{undulationIndex}, \gls{ellipticityIndex}, \gls{nonsphericityIndex},
    \gls{aspectRatio}, and \gls{l2NormMeanCurvature} were significantly
    different between their ruptured and unruptured groups --- we found the
    same in our samples except for the aspect ratio, which was not
    significantly different between these ruptured and unruptured groups. The
    authors concluded that shape is probably more effective than size alone in
    indicating rupture potential.

\begin{figure}[!t]
        \caption{%
            Correlation plots between $(\gls{maxStretch})_{\gls{maximum}}$ in
            \gsupsub{surface}{lumen}{aneurysm} and (a)
            \gls{l2NormGaussianCurvature} and (b) \gls{l2NormMeanCurvature},
            and (c) between the
            $(\gls{maxPrincipalCauchyStress})_{\gls{maximum}}$ and maximum dome
            diameter, \gls{domeDiameter}. Pearson's correlation coefficient and
            the p-values are shown above each plot. The regression lines were
            computed with linear regression using Python's library SciPy.
        }

        \includegraphics[%
            width=\textwidth
        ]   {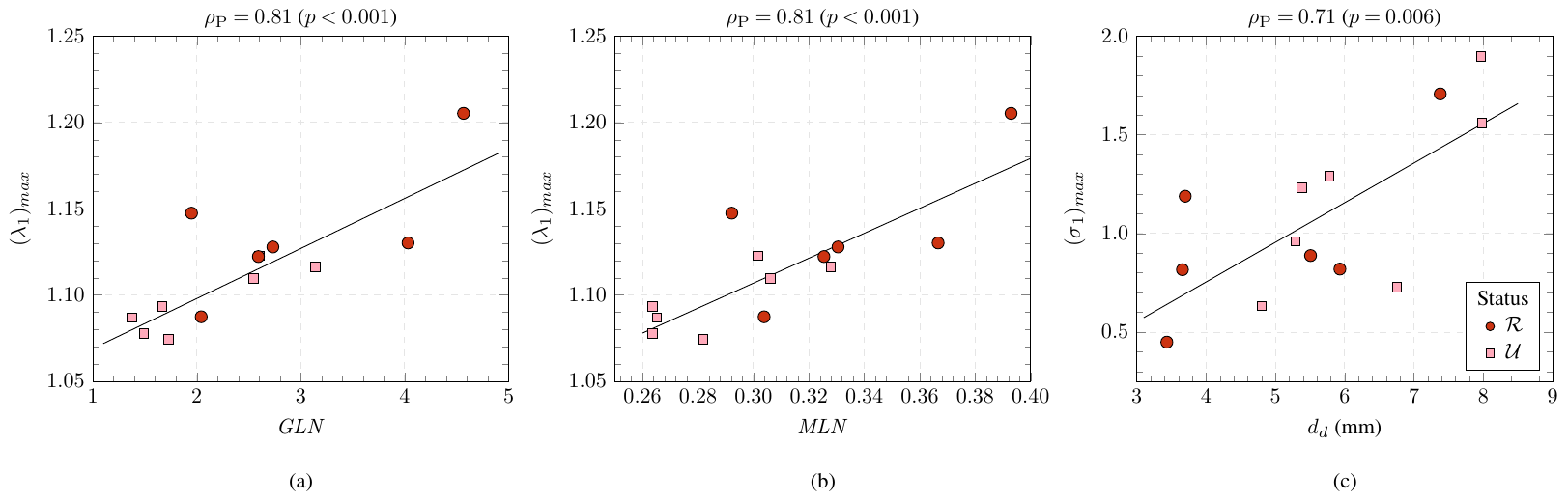}

        \label{fig:significantCorrMechVsSacMetrics}
    \end{figure}

    Nevertheless, we did not find assessments of the correlations between these
    curvature-based metrics --- or even the shape ones --- and the stress and
    stretch of \glspl{ia} walls. Thus, we investigated the relationship between
    all the size, shape, and curvature-based metrics in \cref{tab:meanIndices}
    and the maximum \gls{maxPrincipalCauchyStress} and \gls{maxStretch} on the
    aneurysm sac lumen, indicated by
    $(\gls{maxPrincipalCauchyStress})_{\gls{maximum}}$ and
    $(\gls{maxStretch})_{\gls{maximum}}$, respectively, (we used the absolute
    maximum because this is the metric that would most probably indicate
    rupture danger). We found statistically significant correlations only
    between $(\gls{maxStretch})_{\gls{maximum}}$ and both
    \gls{l2NormGaussianCurvature} and \gls{l2NormMeanCurvature}, and between
    $(\gls{maxPrincipalCauchyStress})_{\gls{maximum}}$ and the maximum dome
    diameter, \gls{domeDiameter} (see
    \cref{fig:significantCorrMechVsSacMetrics}). Therefore, although previous
    works and the current one found that shape indices --- such as
    \gls{undulationIndex}, \gls{nonsphericityIndex}, and \gls{ellipticityIndex}
    --- statistically identified ruptured aneurysms, the actual stretches were
    significantly correlated with the curvature-based metrics only. The same
    was not the case for the stress, although interestingly,
    $(\gls{maxPrincipalCauchyStress})_{\gls{maximum}}$ increased with the
    overall size of the aneurysm. Therefore, curvature-based metrics are more
    likely to bridge the gap between the geometry of a patient-specific
    \gls{ia} and  its wall mechanics, hence potentially acting as more reliable
    indicators of rupture.

    The correlation between stretch and \gls{l2NormMeanCurvature} and
    \gls{l2NormGaussianCurvature} can be explained by the high stretch levels
    found in hyperbolic patches and thus connected to the correlations found
    between surface-average stretch and the Gaussian and mean curvatures (see
    \cref{fig:corrMechanicalVarVsCurvatures}c and d in the online material).
    Larger areas of the sac with negative and high, in absolute terms, Gaussian
    curvature tend to increase \gls{l2NormGaussianCurvature} and, mechanically,
    produce large stretch levels. Although it is important to highlight that
    significant correlations were only found for the maximum over the \gls{ia}
    lumen and not other metrics (we tested the sac surface average, a simple
    average, and the \nth{99} percentile over the sac). Nevertheless, this
    finding further reinforces the potential use of curvature-based metrics for
    rupture assessment. The lack of significant correlations between the
    stretch levels and the other shape metrics should be further investigated.
    It could be that the shape metrics, like the undulation index, are not
    directly influenced by the hyperbolic patches as the curvature-based ones
    are, thus not leading to significant correlations with the maximum stretch.

    Interestingly enough was the increase in
    $\gls{maxPrincipalCauchyStress}_{\gls{maximum}}$ with the maximum diameter.
    This relationship should also be further explored in more extensive studies
    once the mechanisms behind failure are further understood. It could also
    indicate an \gls{ia} near a ruptured state. Nevertheless, we also compared
    the maximum stress and stretch by rupture-status groups and, actually, the
    stress was on average higher in the group \gls{unrupturedGroup}. In
    contrast, only the maximum stretch correctly identified the ruptured
    aneurysms.

    Finally, if sac metrics are to be used to predict the likelihood of
    \gls{ia} rupture, the curvature-based ones, more specifically,
    \gls{l2NormMeanCurvature} and \gls{l2NormGaussianCurvature}, should be the
    subjects of further studies in the future. They could be added, for
    example, to the group of metrics used in investigations that search for a
    score that would predict the likelihood of rupture
    \citep{RajabzadehOghaz2020a}. Other metrics could also be proposed that
    more directly reflect the more significant levels of stress and stretch on
    highly-curved hyperbolic patches.

    \subsection*{Limitations}

    Regarding the material law employed for the \gls{ia} tissue, currently
    there is no agreement on the most correct law to be used for \gls{ia}
    tissue; hence we employed the \gls{mr} law due to the availability of its
    materials constants for this tissue type. For completeness, we also
    performed the simulations with the Yeoh and isotropic Fung-like
    hyperelastic laws and found the same conclusions \citep{Oliveira2022}.
    Additionally, it is important to note that we have used the isotropic
    version of \gls{mr}, despite the tissue of arteries and \glspl{ia} being
    anisotropic, based on evidence that an anisotropic law yielded similar
    mechanical responses compared to its isotropic version
    \citep{Ramachandran2012}.

    Regarding the wall motion modelling, an important feature to be included
    would be the \gls{pae}, which we could not add to the numerical model due
    to the lack of information in the patient-specific images. It is known that
    the \gls{pae} may influence the sac shape \citep{Ruiz2006} and,
    consequently, stress and stretch there. Therefore, future studies should
    include it in their modelling to assess how it will likely impact the
    stress and strain fields.

    An important limitation was to use of a \gls{1wfsi} numerical strategy,
    which ignores the full coupling between the fluid and solid domains.
    First, it is essential to note that this was a reasonable choice of solver
    because we were not concerned with the changes induced in the local
    hemodynamics by the wall motion. On the contrary, in this study, only the
    wall motion was important; therefore, we judged it sufficient to simulate
    the wall motion by imposing realistic flow forces in each geometry.  In any
    case, it is natural to ask how much a \gls{2wfsi} could influence the main
    results obtained here, i.e. the fields of stretch and stress. The main
    force that determines the stresses and stretches of the vascular and
    aneurysmal walls is caused by transmural pressure. Thus, we could verify by
    how much the luminal pressure field would differ, at the peak systole, if
    the fluid domain had the configuration of the vascular wall predicted with
    the \gls{1wfsi} simulations at the peak-systole instant. The results of a
    \gls{cfd} simulation with the vascular geometry at that instant (we
    employed the geometries of the cases \case{case4unruptured} and
    \case{case4ruptured}) showed that the pressure field was almost unchanged,
    qualitatively. Its metrics on the aneurysm surface were affected by less
    than \SI{1}{\percent} in the two aneurysms tested. Therefore, this suggests
    that the stress and stretch fields, and consequently their metrics, would
    not change considerably to the point of altering our main conclusions.

    Finally, part of the conclusions of this work depend on the specification
    of the thresholds defining the abnormal-hemodynamics patches. It is
    important to note that these were defined based on \emph{a priori}
    \gls{cfd} simulations performed with the rigid-wall model and not on the
    deformed configuration of the wall. That was not possible due to the
    \gls{1wfsi} numerical strategy used, which does not deform the fluid
    domain. Nonetheless, the abnormal-hemodynamics patches were built based on
    temporal-averaged quantities (\gls{tawss} and \gls{osi}). Hence, it is
    unlikely that the artery deformation would substantially change these
    patches. Additionally and most importantly, the connection between local
    curvature and local hemodynamics was further reinforced by the data
    provided in \cref{fig:boxPlotsLocalHemodynamics} that was independent of
    those threshold values.

\section{Conclusions} \label{sec:conclusions}

    The local mechanical response of patient-specific \gls{ia} walls has been
    the subject of few studies, which is a surprise given that such an analysis
    could further help explain their rupture. Our study filled this gap by
    providing data on the stress and stretch fields from numerical simulations
    and investigating their levels on different patches of the aneurysm sacs.
    The results revealed that a clear relationship between stress and stretch
    and curvature exists and, moreover, that the curvature of the sac wall
    indicates the regions where the largest stress and stretch occurs. This
    relationship may help in faster predictions of those mechanical variables
    in \glspl{ia} walls without resorting to numerical simulations, which may
    take days. Finally, our findings could potentially guide future studies
    into a better understanding of the rupture event, and suggestions were
    proposed for metrics to be used in that endeavour that would bridge the gap
    between physics and medical practice.

    \section{Acknowledgements}

    We are more than thankful to Prof. Dr. Julio Militzer (in memoriam),
    Dalhousie University, who participated actively in all the steps to
    accomplish this work. This research was supported by resources supplied by
    the Center for Scientific Computing (NCC/GridUNESP) of the \gls{unesp}
    (\url{www2.unesp.br/portal#!/gridunesp}), by ACENET (\url{www.ace-net.ca})
    through Dalhousie University and Compute Canada
    (\url{www.computecanada.ca}).  Funding: This research was supported by
    grants 2017/18514-1 and 2019/19098-7 of the \gls{fapesp}.

    \bibliographystyle{etc/elsarticle-num-names}
    \bibliography{references}

    \appendix
    \newpage
\pagestyle{empty}
%\fancyfoot{}
%\fancyhead{}
\counterwithin{equation}{section}
\renewcommand{\thefigure}{S\arabic{figure}}
\setcounter{figure}{0}

\section*{Supplementary Material} \label{sec:supplementaryMaterial}

\begin{figure}[!htb]
        \caption{%
            Correlation plots between \savg{\gls{maxPrincipalCauchyStress}} (a
            and b) and \savg{\gls{maxStretch}} (c and d) \emph{versus}
            \savg{\gls{GaussianCurvature}} and \savg{\gls{meanCurvature}} over
            the surface patches of elliptic and hyperbolic types, segregated by
            rupture-status groups. The hyperbolic and elliptic \emph{regions}
            on each plot are marked as different colours. The regression lines
            were computed with linear regression.
        }

        \includegraphics[%
            width=\textwidth
        ]   {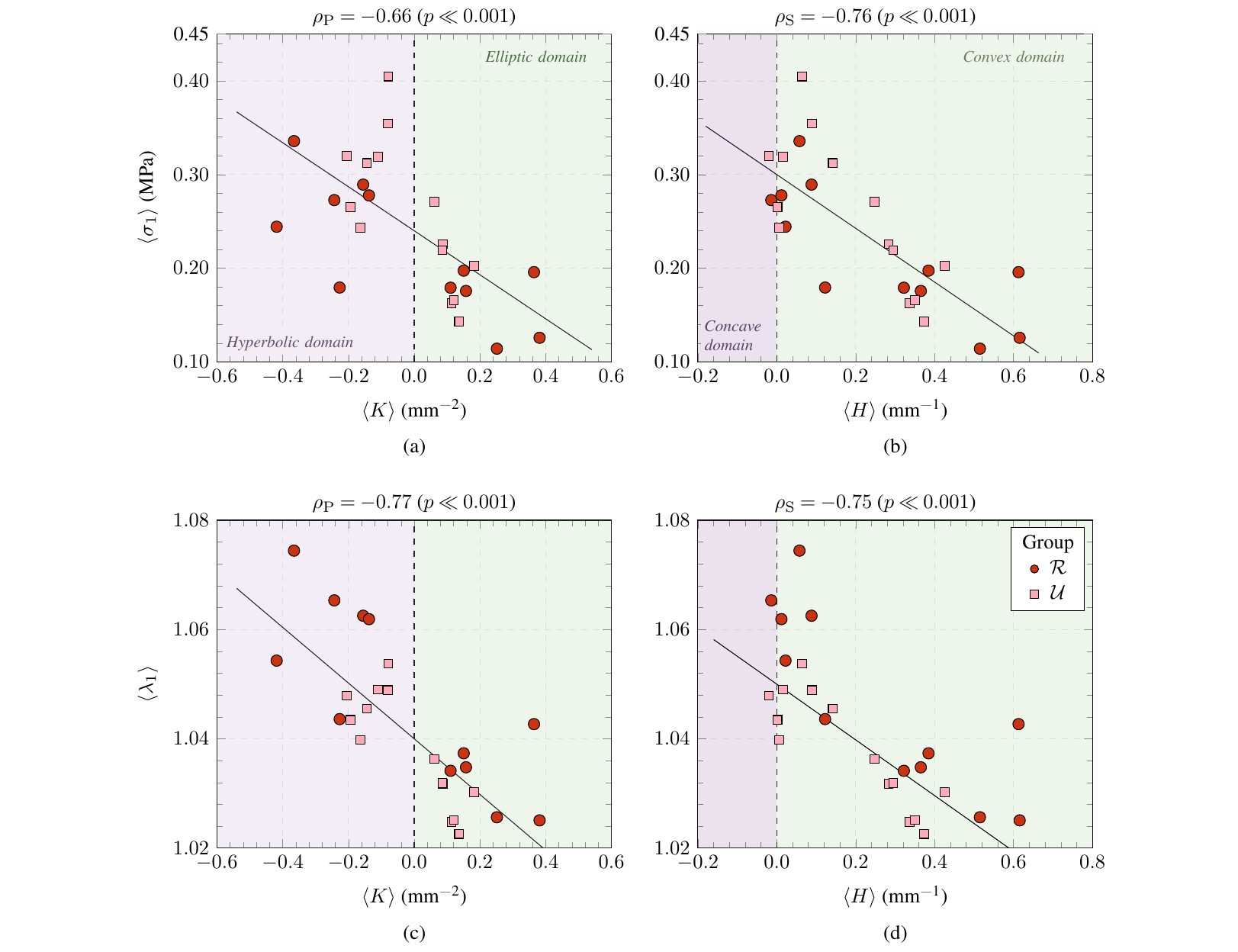}

        \label{fig:corrMechanicalVarVsCurvatures}
    \end{figure}

    To further support the relationship found between stress and stretch and
    the local shape of the aneurysm sac surfaces, we plotted
    \savg{\gls{maxPrincipalCauchyStress}} and \savg{\gls{maxStretch}} against
    the surface average of the curvatures on each local-shape patch type
    (surface-averages of the curvatures were also used due to the noise in the
    curvature fields). The data points and the regression lines, computed using
    Python's library SciPy, are shown in
    \cref{fig:corrMechanicalVarVsCurvatures}.

\end{document}